\documentclass[conference]{IEEEtran}
\IEEEoverridecommandlockouts

\usepackage{cite}
\usepackage{amsmath,amssymb,amsfonts}
\usepackage{textcomp}
\usepackage{xcolor}
\usepackage{graphicx}
\usepackage{amsmath,amssymb,amsfonts}
\usepackage{textcomp}
\usepackage{xcolor}
\usepackage{algorithm}
\usepackage{algorithmic}
\usepackage{multirow}
\usepackage{multicol}
\usepackage{booktabs}
\usepackage{threeparttable}
\usepackage{amsmath}

\newtheorem{theorem}{Theorem}
\newtheorem{definition}{Definition}
\newtheorem{lemma}{Lemma}

\usepackage{subfigure}
\usepackage{setspace}
\usepackage{epstopdf}

\def\BibTeX{{\rm B\kern-.05em{\sc i\kern-.025em b}\kern-.08em
    T\kern-.1667em\lower.7ex\hbox{E}\kern-.125emX}}
\begin{document}

\title{A Crypto-Assisted Approach for Publishing Graph Statistics with Node Local Differential Privacy}


\author{\IEEEauthorblockN{Shang Liu}
\IEEEauthorblockA{
\textit{Kyoto University}\\
Kyoto, Japan \\
shang@db.soc.i.kyoto-u.ac.jp}
\and
\IEEEauthorblockN{Yang Cao}
\IEEEauthorblockA{
\textit{Hokkaido University}\\
Sapporo, Japan \\
yang@ist.hokudai.ac.jp}
\and
\IEEEauthorblockN{Takao Murakami}
\IEEEauthorblockA{
\textit{AIST}\\
Tokyo, Japan \\
takao-murakami@aist.go.jp}
\and
\IEEEauthorblockN{Masatoshi Yoshikawa}
\IEEEauthorblockA{
\textit{Kyoto University}\\
Kyoto, Japan \\
yoshikawa@i.kyoto-u.ac.jp}
}
\maketitle

\begin{abstract}
	Publishing graph statistics under node differential privacy has attracted much attention since it provides a stronger privacy guarantee than edge differential privacy. 
	Existing works related to node differential privacy assume a trusted server who holds the whole graph. 
	However, in many applications, a trusted curator is usually not available due to privacy and security issues.
	In this paper, for the first time, we investigate the problem of publishing graph statistics under \emph{Node Local Differential privacy (Node-LDP)}, which does not rely on a trusted server.
    We propose an algorithm to publish the degree distribution with Node-LDP by exploring how to select the graph projection parameter in the local setting and how to execute the graph projection locally. 
    Specifically, we propose a crypto-assisted local projection method based on cryptographic primitives, achieving the higher accuracy than our baseline pureLDP local projection method.
    Furthermore, we improve our baseline graph projection method from node-level to edge-level that preserves more neighboring information, owning better utility.
	Finally, extensive experiments on real-world graphs show that crypto-assisted parameter selection owns better utility than pureLDP parameter selection, 
	and	edge-level local projection provides higher accuracy than node-level local projection, improving by up to 57.2\% and 79.8\%, respectively.
	
\end{abstract}


\section{Introduction}
Graph analysis has been receiving more and more attention on social networks, transportation, protein forecast, etc. 
However, directly publishing graph statistics may leak sensitive information about an individual~\cite{RHMS_2}. 
Recently, many research works have studied the problem of publishing sensitive graph statistics under differential privacy (DP) \cite{dwork_foundations,li2016differential}.
Compared with previous privacy models (e.g., $k$-anonymity, $l$-diversity, $t$-closeness), differential privacy can resist most private attacks and provide a provable privacy guarantee. 

When DP is applied to graph analysis, there are two common variants of DP \cite{task2012guide,li2021private}: Edge Differential Privacy~\cite{qian2018publishing, hay2009accurate, karwa2012differentially, proserpio2012workflow} and Node Differential Privacy \cite{day2016publishing, kasiviswanathan2013analyzing, raskhodnikova2016lipschitz, blocki2013differentially}. 
Intuitively, Edge Differential Privacy guarantees that a query result does not significantly reveal sensitive information about a particular edge in a graph, while Node Differential Privacy protects the information about a node and all its adjacent edges. 
Obviously, Node Differential Privacy provides a much stronger privacy guarantee than Edge Differential Privacy. 
Existing works related to Node Differential Privacy are almost in the central (or global) model, where a trusted curator holds the entire graph data before data publishing.
We refer to the above two variants under a central server setting as Edge Central Differential Privacy (Edge-CDP) and Node Central Differential Privacy (Node-CDP), respectively.
However, the assumption about a trusted server may not be practical in many applications (i.e., individual contact lists) due to security reasons, such as privacy leaks and breaches in recent years \cite{yang2020local}. 
Local differential privacy (LDP) \cite{duchi2013local, kasiviswanathan2011can} is a promising model that does not require a trusted server to collect user information.
In LDP, each user perturbs its sensitive information by herself and sends perturbed messages to the untrusted server; hence it is difficult for the curator to infer sensitive information with high confidence.
We refer to the above two variants of DP without a trusted server as Edge Local Differential Privacy (Edge-LDP) and Node Local Differential Privacy (Node-LDP), respectively.

Although there are many recent studies on publishing statistics under Edge-LDP \cite{AsgLDP, qin2017generating, imola2021locally}, to the best of our knowledge, no existing work in literature attempts to investigate graph statistics release under Node-LDP.
Basically, it is very challenging to publish graph statistics under Node-LDP due to the lack of global view and prior knowledge about the entire graph. 
Consider querying the node degree in a social graph, and if two graphs differ in one node, the results may differ at most $(n-1)$ edges in the worst case, where $n$ is the number of users.
Thus the sensitivity of Node Differential Privacy is $O(n)$ while that of Edge-DP is $O(1)$.
Naively scaling the sensitivity of Edge-LDP for achieving Node-LDP suffers the prohibitive utility drop.

Graph projection \cite{kasiviswanathan2013analyzing, blocki2013differentially, day2016publishing} is the key technique to reduce the high sensitivity, but existing projections are only designed for the central model.
When attempting to apply central graph projections into Node-LDP, it is difficult for each local user to project its neighboring information with a limited local view.
In central models, with the global view, the server can determine optimal strategies of removing which edges or nodes to maximize the overall utility.
However, in the local setting, each user can only see its own information but not other neighboring information. 
What's more, it is difficult for local users to obtain a graph projection parameter $\theta$ with high accuracy as they have little knowledge about the entire graph.
In general, graph projection transforms a graph into a $\theta$-bounded graph whose maximum degree is no more than $\theta$.
The parameter $\theta$ plays a vital role as it reduces the sensitivity from $O(n)$ to $O(\theta)$.
If $\theta$ is too small, a large number of edges will be removed during the projection. 
If $\theta$ is too large, the sensitivity will become higher and more noise will be added during the protection.  
Graph projections in the central setting can easily opt for the desirable projection parameter $\theta$ with some prior knowledge of the whole graph, for instance, the maximum degree, the average degree, etc.; yet it is harder for local users to achieve it, since they have little prior knowledge about the entire graph.

\begin{figure}[t]
	\centering  
	\includegraphics[width=0.95\linewidth]{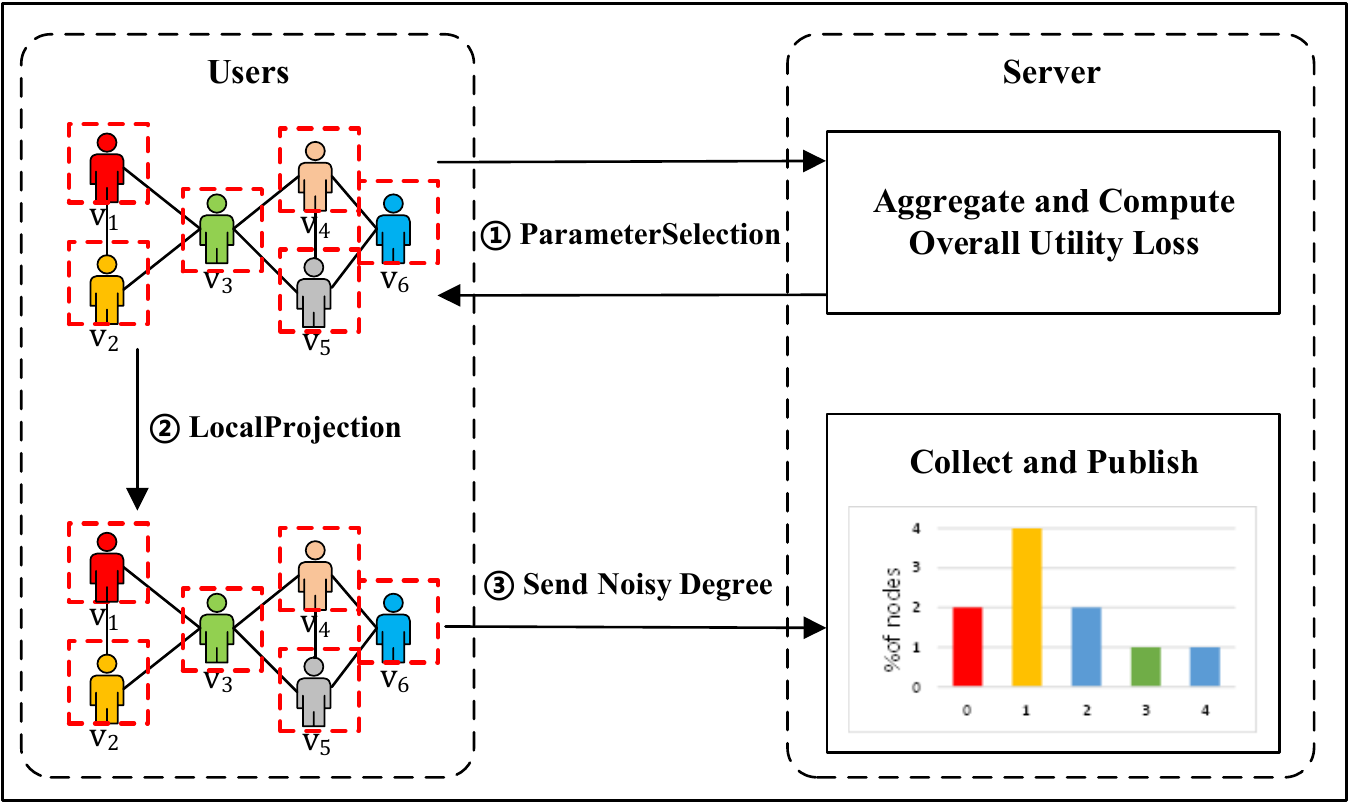}
	\caption{Framework of our methods}  
	\label{fig:framework}  
\end{figure}

In this paper, we introduce a novel local graph projection method for publishing the degree distribution under Node-LDP by addressing two main challenges:
(1) How to obtain the graph projection parameter $\theta$ in the local setting; 
(2) How to execute the graph projection locally.
The general framework is depicted in detail in Fig. \ref{fig:framework}, which includes three phases: 
(1) local users and server collaboratively select a projection parameter $\theta$ with minimum utility loss (Sec.\ref{sec:projection_parameter}); 
(2) local users execute local graph projection based on selected parameters (Sec.\ref{sec:local_projection});
(3) local users perturb individual information and send noisy degrees to the server.

First, to find the optimal projection parameter $\theta$, we design a multiple-round protocol to find which parameter has the minimum utility loss. 
Specifically, for each round, each user calculates the potential utility loss with respect to a certain $\theta$ and sends to the server for computing the aggregated loss. 
The utility loss contains sensitive information since it is calculated based on each user’s raw data.
We design two methods to protect individual messages based on different privacy-enhancing techniques: pureLDP and crypto-assisted.
The pureLDP method is a naive local graph projection method under Node-LDP.
The obvious disadvantage is that multiple-round adding noise significantly deteriorates the utility.
To improve it, we design a crypto-assisted parameter selection method that improves the utility with cryptographic primitives.
The key challenge is that aggregated utility loss is computed for the evaluation while individual utility loss can be protected.
We first use the order-preserving encryption (OPE) scheme~\cite{kamara2020review,tueno2020efficient} to encode individual information for comparing different utility loss values.
Then, we mask the encrypted value with Secure Aggregation (SA) technique \cite{bonawitz2017practical} to protect the order information of individual utility loss.
The masks can be cancelled during the aggregation and the final aggregated utility loss is protected under OPE scheme.

Second, we propose two different local projection methods based on different granularity, including node-level method and edge-level method.
In node-level method, each node is the minimal unit of a graph and correlations among neighboring users will be ignored coarsely.
However, this approach loses too much neighboring information that significantly influences the overall utility (detailed analysis in Sec. \ref{sec:node-level}). 
Then, we propose an improved approach, edge-level method, where each edge is the minimal unit that is more fine-granularity information.
One main challenge is that privacy leakage may happen via communication messages among neighboring users.
We represent this message as an operation vector and carefully design a randomized mechanism to perturb each bit of this vector while satisfying Node-LDP.
As a result, it is difficult for neighboring users to distinguish whether the current node degree is larger than $\theta$ or smaller than $\theta$.

Our contributions can be summarized as follows:
\begin{itemize}
	\item We propose and study the problem of publishing the degree distribution under Node-LDP for the first time. 
	We give a detailed description of the problem definition and conclude the research gap. 
	We present an overview of publishing the degree distribution under Node-LDP.
	
	\item We design two methods to select the projection parameter $\theta$ in the local setting: pureLDP and crypto-assisted. 
    Crypto-assisted method guarantees the security of individual utility loss with cryptographic primitives, which achieves a higher accuracy than the baseline pureLDP method.
    
	\item We design two local graph projection approaches based on different granularity: node-level and edge-level. 
	The improved edge-level method preserves more information and provides better utility than the baseline node-level method.

	\item Extensive experiments on real-world graph datasets validate the correctness of our theoretical analysis and the effectiveness of our proposed methods.
\end{itemize}

\section{Problem Definition and Preliminaries}
\label{sec:preliminaries}
\subsection{Problem Definition}
\label{sec:problem}
\begin{figure}[t]
	\centering  
	\includegraphics[width=\linewidth]{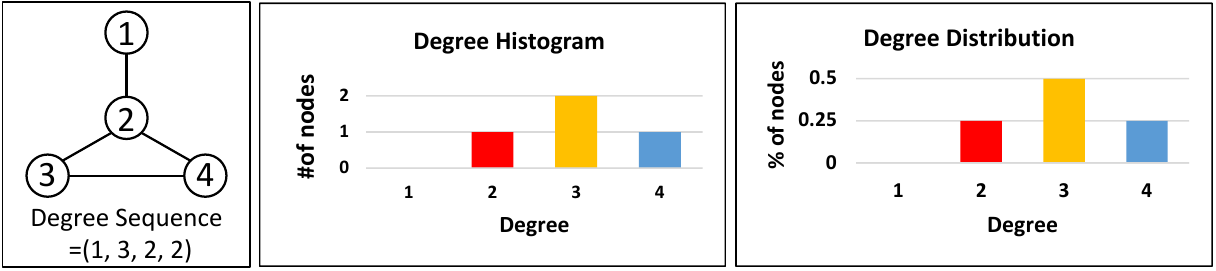}
	\caption{Example of publishing degree distribution}  
 	\label{fig:degree_example}  
\end{figure}

In this paper, we consider an undirected graph with no additional attributes on nodes or edges. 
An input graph is defined as $G=(V,E)$, where $V=\{v_1,...,v_n\}$ is the set of nodes, where $|V|=n$, and $E\subseteq V \times V$ is the set of edges. 
For each user $i$, $B_i=\{b_{i1},b_{i2},...,b_{in}\}$ is its adjacent bit vector, where $b_{ij}=1$ if the edge $(v_i,v_j)\in E$ and $b_{ij}=0$ otherwise. 
The number of adjacent edges for one node $i$ is the node degree $d_i$, namely, $d_i=\sum_{j=1}^n b_{ij}$. 
The server collects a perturbed degree sequence $seq=\{d_1, d_2,..., d_n\}$ from each local user and publishes the degree histogram $hist(G)$. 
The degree distribution $dist(G)$ can be easily obtained from $hist(G)$ by counting each degree frequency.
Fig. \ref{fig:degree_example} shows an example of degree sequence, degree histogram, and degree distribution, respectively.

We use two common measures to assess the accuracy of our algorithms. First, we use the mean squared error (MSE)~\cite{LF-GDPR} to estimate the error between noisy histogram ${hist(G)}^\prime$ and original histogram $hist(G)$. Generally, the MSE can be computed as MSE$(hist(G), hist(G)^\prime)=\frac{1}{n}\sum_{i=1}^n(hist(G)_i- {hist(G)_i}^\prime)^2$, where $n$ is the number of users in a graph. Also, we compute the mean absolute error (MAE)\cite{willmott2005advantages} which can be represented by MAE$(hist(G), hist(G)^\prime)=\frac{1}{n}\sum_{i=1}^n|hist(G)_i- {hist(G)_i}^\prime)|$.

\subsection{Preliminaries}
\if.
According to different trusted assumptions, differential privacy can be divided into two types: central differential privacy (CDP) in Definition \ref{def:CDP} and local differential privacy (LDP) in Definition \ref{def:LDP}. 
\begin{definition}[CDP]
	\label{def:CDP}
	A random algorithm $M$: $\mathbb{X}$$^n$ $\rightarrow$ $\mathbb{Z}$ satisfies $\epsilon$-DP, where $\epsilon$ $\geq$ 0, iff for any two neighboring datasets D, $D^\prime$ $\in$ $\mathbb{X}$$^n$, any subsets S $\subseteq$ $\mathbb{Z}$,  
	\begin{center}
		$Pr[M(D) \in S] \leq e^{\epsilon} Pr[M(D^{\prime}) \in S]$
	\end{center}  
\end{definition}

\begin{definition}[LDP]
	\label{def:LDP}
	A random algorithm $M$: $\mathbb{X}$ $\rightarrow$ $\mathbb{Y}$ satisfies $\epsilon$-LDP, where $\epsilon$ $\geq$ 0, iff for any two input x, x$^{\prime}$ $\in$ $\mathbb{X}$, any output y $\in$ $\mathbb{Y}$,  
	\begin{center}
		$Pr[M(x) = y] \leq e^{\epsilon} Pr[M(x^{\prime}) = y]$
	\end{center}  
\end{definition}
\fi

Since the trusted third party is impractical, LDP has become the de facto standard of privacy protection to protect individual information. 
As a graph consists of nodes and edges, there are two definitions when LDP is applied to either of them: edge local differential privacy (Edge-LDP) in Definition \ref{def:edgeLDP} and node local differential privacy (Node-LDP) in Definition \ref{def:nodeLDP}.

\begin{definition}[Edge-LDP]
	\label{def:edgeLDP}
	A random algorithm $M$ satisfies $\epsilon$-Edge-LDP, iff for any $i \in [n]$, two adjacent bit vectors $B_i$ and \textbf{$B_i^{\prime}$} that differ only one bit, and any output y $\in range(M)$, 
	\begin{center}
		$Pr[M(B_i) = y] \leq e^{\epsilon} Pr[M(B_i^{\prime}) = y]$
	\end{center} 
\end{definition}

\begin{definition}[Node-LDP]
	\label{def:nodeLDP}
	A random algorithm $M$ satisfies $\varepsilon$-Node-LDP, iff for any $i \in [n]$, two adjacent bit vectors $B_i$ and \textbf{$B_i^{\prime}$} that differ at most $n$ bits, and any output $y\in range(M)$,  
	\begin{center}
		$Pr[M(B_i) = y] \leq e^{\varepsilon} Pr[M(B_i^{\prime}) = y]$
	\end{center}
\end{definition}

Node-LDP is clearly a much stronger privacy guarantee than Edge-LDP since it requires hiding the existence of each node along with its incident edges. 
To our knowledge, however, there are few research works that release graph statistics under Node-LDP. Although Zhang $et$ $al.$\cite{zhang2020differentially} consider Node-DP in the local setting where each node represents a software component and an edge represents control flow between components, the directed graphs on the control-flow behavior of different users are mutually independent. 
We consider a totally different setting where each node represents a user and each edge represents the correlation between neighboring users. 

There are two kinds of DP, namely, $bounded$ DP and $unbounded$ DP \cite{li2016differential, 10.1145/1989323.1989345}. 
In a bounded DP, two neighboring datasets $D$, $D'$ have the same size $n$ and $D\prime$ is obtained from $D$ by changing or replacing one element. 
In unbounded DP, $D\prime$ can be derived from $D$ by deleting or adding one element. Here, we use the bounded DP to publish the degree distribution. 
That is to say, the size of each adjacent bit vector is equal to $n$, where $n$ is the number of users. 
Node-LDP satisfies the post-processing property (Theorem \ref{theorem:post-process}) and the composition property (Theorem \ref{theorem:composition}) \cite{dwork_foundations}. 

\begin{theorem}[Post-Processing]
	\label{theorem:post-process}
	If a randomized algorithm $R$ satisfies $\varepsilon$-DP, then for an arbitrary randomized algorithm $S$, $S \circ R$ also satisfies $\varepsilon$-DP.
\end{theorem}

\begin{theorem}[Composition Property]
	\label{theorem:composition}
	$\forall \varepsilon \geq 0, k\in N$, the family of $\varepsilon$-DP mechanism satisfies $t\varepsilon-DP$ under t-fold adaptive composition.
\end{theorem}

To satisfy DP, one way to add some noise into the query result. 
In the Laplace mechanism (Theorem \ref{theorem:laplace}) \cite{dwork_foundations, li2016differential}, given the privacy budget $\varepsilon$ and sensitivity $\triangle$, one publishes the result after adding Lap($\frac{\triangle}{\epsilon}$) noise.    
\begin{theorem}[Laplace Mechanism]
	\label{theorem:laplace}
	For any function $f$, the Laplace mechanism $A(D)=f(D)+Lap$ ($\frac{\triangle f}{\varepsilon}$) satisfies $\varepsilon$-DP.
\end{theorem}

\if
Additionally, the randomized response (RR) \cite{warner1965randomized, erlingsson2014rappor} is a common method to guarantee LDP. In specific, each user gives the true answer with the flipping probability $p$ and the opposite answer with probability $1-p$. Some works have proved that RR satisfies $\varepsilon$-LDP if $p=\frac{e^\varepsilon}{1+e^\varepsilon}$ \cite{LF-GDPR, AsgLDP}.
\fi

\section{Overview of Proposed Methods}
\label{sec:overview}
We aim to design a method for publishing the degree distribution that approximates the original distribution as possible while satisfying the strict Node-LDP. Our proposed methods support the following functions: 
1) obtaining the graph projection parameter $\theta$ in the local setting; 
2) conducting the graph projection locally; 
3) publishing the degree distribution under Node-LDP.

\begin{algorithm}[h]
	\caption{Publishing the degree distribution} 
	\label{alg:degree_distribution} 
	\begin{algorithmic}[1] 
		\REQUIRE  
		Adjacent bit vectors $\{B_1,..., B_n\}$, \\
		\qquad privacy budget $\varepsilon_1, \varepsilon_2, \varepsilon_3$
		\ENSURE  
		A noisy degree distribution $dist(G)^\prime$ 
		
		\STATE $\theta \leftarrow \mathtt{SelectParameter}$($\{B_1,...,B_n\}, \varepsilon_1$) // Sec. \ref{sec:projection_parameter} \\
		/* User side. \qquad \qquad   */
		\FOR{each user $i\in \{1,2,...,n\}$}
		\STATE $\hat{d_i} \leftarrow \mathtt{LocalProjection}$($B_i$, $\theta, \varepsilon_2$) // Sec. \ref{sec:local_projection}
		\STATE $d_i^\prime \leftarrow \hat{d_i}$+Lap($\frac{2\theta}{\varepsilon_3}$)
		\STATE User $i$ sends $d_i^\prime$ to server
		\ENDFOR 
		/* Curator side. \qquad \qquad  */
		\STATE Curator collects all noisy degree $d_i^\prime$
		\RETURN $dist(G)^\prime$
	\end{algorithmic} 
\end{algorithm}

We provide an overview of our solutions in Algorithm~\ref{alg:degree_distribution}. 
First, a private parameter selection method is designed to select the projection parameter with minimum utility loss in the local setting (Section \ref{sec:projection_parameter}).
The curator collects individual utility loss from local users and evaluates each candidate projection parameter $k$ by computing the aggregated utility loss.
To protect sensitive individual utility loss during communications, we first propose one naive approach, pureLDP parameter selection, which adds noise into individual utility loss. 
However, this method adds too much noise to destroy the order information of different aggregated utility loss, significantly influencing the selection accuracy.
Then, we propose an improved crypto-assisted parameter selection method using cryptographic primitives.
Specifically, the individual utility loss is encrypted by order-preserving encryption (OPE) \cite{popa2013ideal} scheme where the numerical order in the plaintext domain will be preserved in the ciphertext domain.
To prevent leaking the order information of individual utility loss while preserving the order of the aggregated utility loss, we add one mask into encrypted values with Secure Aggregation technique~\cite{bonawitz2017practical}.
The added masks are cancelled after the aggregation and the final aggregated utility loss is protected under OPE scheme.

Second, as soon as the projection parameter is decided, each user can execute the local projection (Section \ref{sec:local_projection}). 
Compared with the Node-CDP, it is more difficult for each user to execute the local projection due to the limited local view of the entire graph.
We first give a baseline node-level approach that is motivated by graph projection \cite{imola2021locally} with Edge-LDP.
In node-level local projection, the node is the minimal unit and correlations among users are ignored. 
It is easy to deploy but lose much information that significantly influences the utility. 
Then we design an improved edge-level local projection where each edge is the minimal unit during the projection. 
The key challenge is that information leakage may happen via mutual edges among neighboring users.
For example, neighboring users may know that the current degree is larger than or less than $\theta$ during the local projection.
We represent this sensitive message as each bit in an operation vector and design a randomized mechanism to perturb each bit.
Thus neighboring users cannot distinguish the current node degree whether larger than $\theta$ or smaller than $\theta$. 

Third, after finishing the local projection, each user perturbs its projected degree using the Laplace mechanism. 
Here, the sensitivity is $2\theta$ since any change of one edge will make an effect on two node degrees. 
Then, they send the noisy degree to the server.
The curator collects the degree sequence and publishes the degree histogram and degree distribution.

\section{Projection Parameter Selection}
\label{sec:projection_parameter}
\subsection{PureLDP Selection}
Intuitively, the server can help local users select the parameter with the minimum utility loss from  the candidate set \{1,2,...,$K$\} through multiple-round communications. 
We design a utility loss function to evaluate each candidate parameter $k$. Our utility loss function has two parts, as shown in Equation \ref{equation:loss_function}, which includes projection utility loss during the local projection and publishing utility loss from added Laplace noise. 
The publishing utility loss $E_D$ is usually a constant value. 
For example, the publishing utility loss of degree distribution is equal to the variance, namely, $E_D=n.2(\frac{2k}{\varepsilon_3})^2=\frac{8nk^2}{\varepsilon_3^2}$.
The projection utility loss $E_P$ is aggregated by all individual projection utility losss, i.e., $E_P=\sum_{i=1}^n\{d_i-k|v_i\in V, d_i>k\}$.
But directly collecting each individual utility loss from local users may reveal personal information.
In baseline method, we use the Laplace mechanism to provide the privacy guarantee and its sensitivity is $(n-1-k)$ in Node-LDP, as shown in Lemma \ref{lemma:sensitivity}. 

\begin{equation}
\label{equation:loss_function}
F(k)=E_P+E_D,
\end{equation}

\begin{center}
$E_P=\sum_{i=1}^n|\{d_i-k|v_i\in V, d_i>k\}|$
\end{center}

\begin{center}
$E_D=n.2(\frac{2k}{\varepsilon_3})^2=\frac{8nk^2}{\varepsilon_3^2}$
\end{center}

\begin{lemma}
	\label{lemma:sensitivity}
	For any projection loss $|d_i-\hat{d_i}|$ and $|d_i-\hat{d_i}|^\prime$, we have 
	\begin{center}
    $||d_i-\hat{d_i}|-|d_i-\hat{d_i}|^\prime|_1 \leq (n-1-k)$
    \end{center}
\end{lemma}

\emph{Proof of Lemma \ref{lemma:sensitivity}:} Given the graph projection parameter is $k$, for each node degree $d_i$, if $d_i \leq k$, projected node degree $\hat{d_i}$ will remain the original value, namely, $\hat{d_i}=d_i$; 
otherwise, $\hat{d_i}=k$. Thus, we have
\begin{equation}
    |d_i-\hat{d_i}|=\left \{
        \begin{array}{lc}
            d_i-\theta, & d_i \textgreater k  \\
            0, & d_i \leq k
         \end{array}
        \right.\nonumber
\end{equation}
Since the maximum node degree is $(n-1)$, the projection loss value is bounded by ($n-1-k$).

\begin{algorithm}[t]
	\caption{PureLDP parameter selection} 
	\label{alg:ldp_enhanced} 
	\begin{algorithmic}[1] 
		\REQUIRE  
		Adjacent bit vectors $\{B_1,...,B_n\}$, privacy budget $\varepsilon_1$
		\ENSURE  
		Projection parameter $\theta$
		\FOR{each integer $k \in \{1,2,...,K\}$}
		\STATE /* User side. \qquad \qquad   */
		\FOR{each user $i\in \{1,2,...,n\}$}
		\STATE $\hat{d_i} \leftarrow$ $\mathtt{LocalProjection}$($B_i$, $k$) // Sec. \ref{sec:local_projection}
		\STATE $d_i \leftarrow \sum_{j=1}^n b_{i,j}$
		\STATE $E_{P_{k,i}} \leftarrow |d_i-\hat{d_i}|$+Lap($\frac{n-1-k}{\varepsilon_1/K}$)
		\STATE User $i$ sends $E_{P_{k,i}}$ to server
		\ENDFOR \\
		/* Curator side. \qquad \qquad   */
		\STATE $E_{P_{k}} \leftarrow \sum_{i=1}^n E_{P_{k,i}}$
		\STATE $\theta \leftarrow k$ when $(E_{P_{k}}+E_D)$ is minimum
		\ENDFOR \\
		\RETURN $\theta$ 
	\end{algorithmic} 
\end{algorithm}

\textbf{Algorithm.} Algorithm \ref{alg:ldp_enhanced} presents the formal description of pureLDP parameter selection. It takes as input a graph $G$ that is represented as bit vectors $\{B_1,...,B_n\}$, the privacy budget $\varepsilon_1$, and the size of candidate parameter $K$. For each candidate parameter $k$, the original graph is first projected to $k$-bounded graph using the local graph projection method (in Section \ref{sec:local_projection}).
Then, each user computes the projection utility loss and adds the Laplace noise into individual utility loss with the sensitivity ($n-1-k$).
After collecting all noisy individual projection utility loss, the server computes the sum of aggregated projection utility loss and publishing utility loss.
Finally, the parameter $\theta$ is selected when the overall utility loss is the minimum and server sends this $\theta$ to each local user.

\begin{figure}[b]
	\centering
	\subfigure[Under $\theta$=20]{
		\begin{minipage}[t]{0.48\linewidth}
			\centering
			\includegraphics[width=\linewidth]{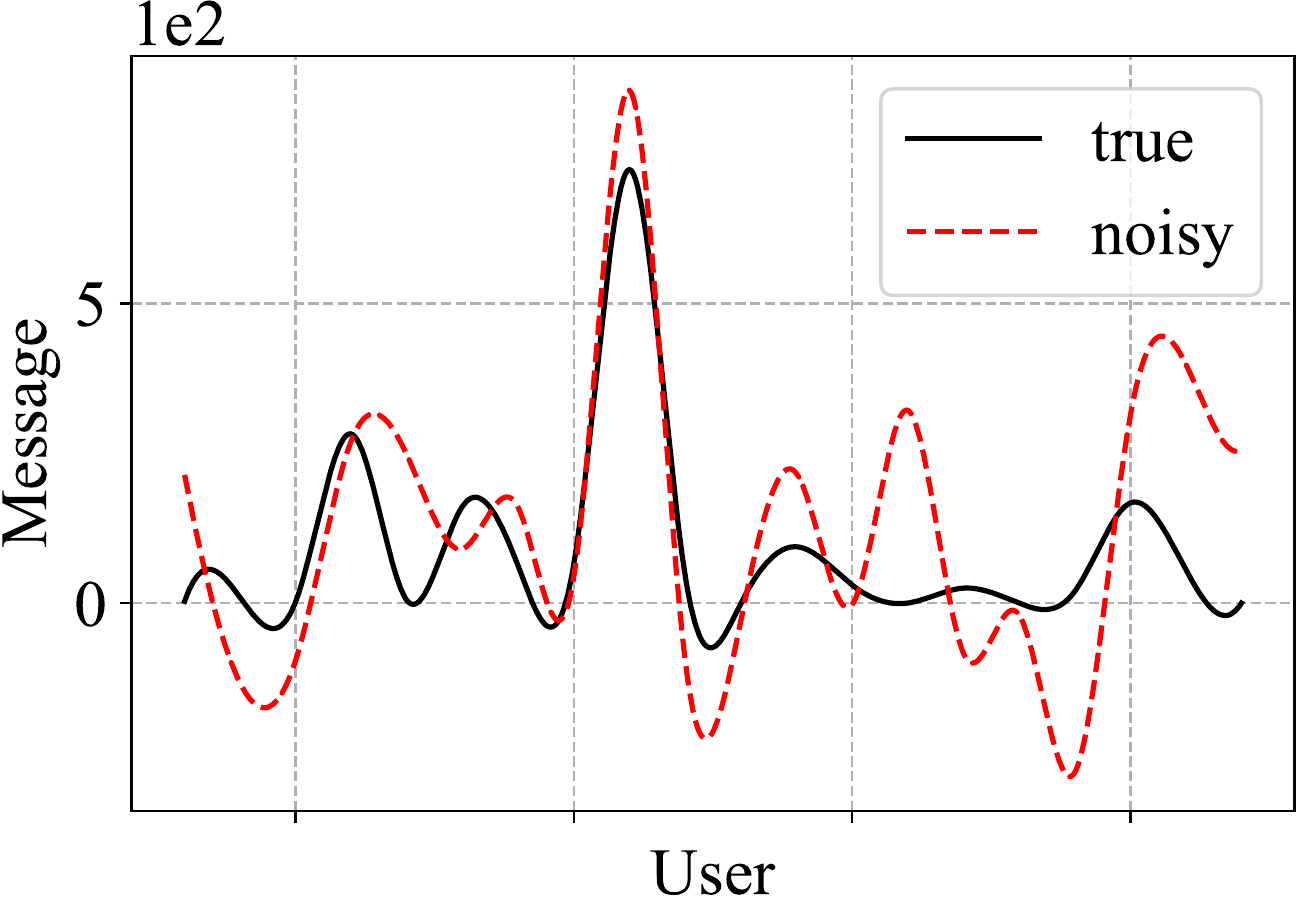}
		\end{minipage}
	}%
	\subfigure[Under various $\theta$]{
		\begin{minipage}[t]{0.45\linewidth}
			\centering
			\includegraphics[width=\linewidth]{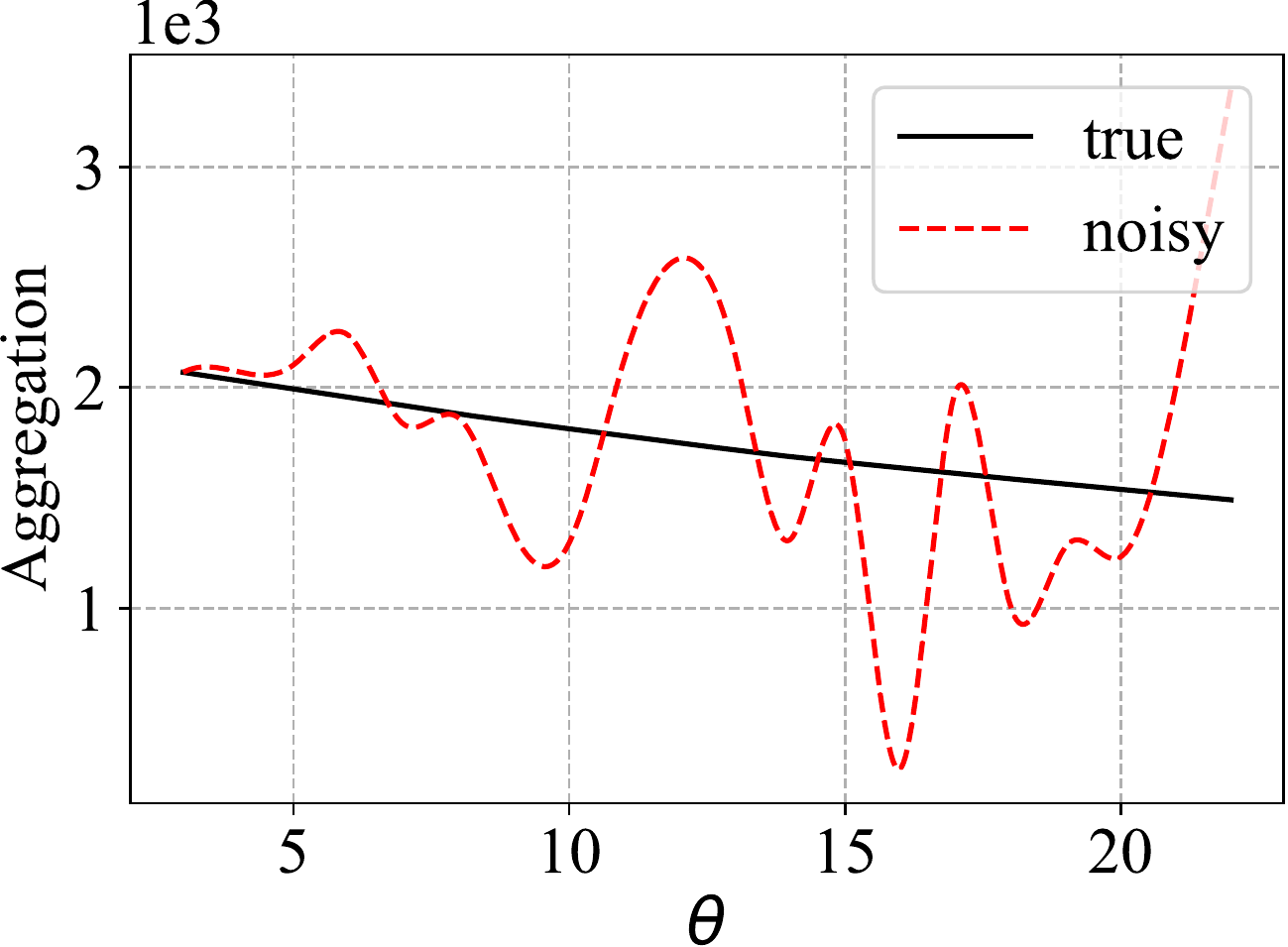}
		\end{minipage}
	}%
\centering
\caption{The impact of added noise on order information}
\label{fig:noise on theta}
\end{figure}

\textbf{Limitation.} 
Much noise is added into the true individual utility loss, which significantly destroys the order information of aggregated utility loss.
To capture the impact of adding Laplace noise on the accuracy of pureLDP parameter selection method, we execute experiments on Wikipedia vote network from SNAP \cite{snapnets}.
As shown in Fig. \ref{fig:noise on theta}, the left figure presents the impact of added noise on the order of individual utility loss when $\theta$ = 20, and the difference between true and noisy utility loss is up to 85\%.
The right one shows the influence on the order of aggregated utility loss under various $\theta$ and the difference is up to 90\%.
Finally, the accuracy of selecting projection parameter $\theta$ is influenced significantly.

\subsection{Crypto-assisted Selection}
Our goal is that each individual projection utility loss can be protected
when the order of aggregated utility loss is preserved.
Order-preserving encryption (OPE) scheme \cite{agrawal2004order, boldyreva2011order} can achieve this idea that 
the $i$-th data in the plaintext domain is transformed to the $i$-th data in the ciphertext domain, so the numerical order among plaintexts is preserved among ciphertexts.
Thus when individual utility loss are encrypted by OPE scheme, the numerical order of individual utility loss can be preserved and the order of aggregated utility loss is also preserved.
But there is the other problem that the order of aggregated utility loss is preserved while the order of individual utility loss is revealed to server.
Next, we use the secure aggregation \cite{bonawitz2017practical} to mask encoded individual utility loss, and these masks can be cancelled during the aggregation.

\textbf{OPE Schemes.}
\label{sec:ope}
There are many existing works related to OPE scheme. 
For example, Popa $et$ $al.$ \cite{popa2013ideal} proposed an interactive OPE scheme between the client and the server, which allows the encrypted state to update over time as the new values are inserted. 
The server organizes the encrypted values by maintaining a binary search tree, namely, OPE tree.
To reduce the high cost of the encryption, Kerschbaum $et$ $al.$ \cite{kerschbaum2014optimal} designed a more efficient OPE scheme that uses a dictionary to keep the state and thus does not need to store too much data.
Roche $et$ $al.$ \cite{roche2016pope} proposed an alternative approach to optimize the heavy insertion of OPE schemes. It is very efficient at insertion and has a lower communication cost, but it provides only a partial order.
Here, we choose a linear OPE scheme \cite{6253544} to encode individual utility loss since it can be directly extended for the local setting.

\textbf{Secure Aggregation}\cite{bonawitz2017practical}.
\label{sec:sa}
Consider a curator with $n$ users where user $i \in [n]$ has its private local vector $x_i$. 
The objective of server is to compute the sum of models $\sum_{i\in n} x_i$ without getting any other information on private local data.
Suppose each pair of users $(i,j), i<j$ agree on some random vector $s_{i,j}$.
If user $i$ adds $s_{i,j}$ to $x_i$ and $j$ subtracts it from $x_j$, then the mask $s_{i,j}$ will be canceled when their vectors are added, but their true inputs will be concealed without revealing.
Formally, each masked value can be computed:
\begin{center}
    $y_i=x_i+\sum\limits_{j\in n:i<j} s_{i,j}-\sum\limits_{j\in n:i>j} s_{i,j}$  (mod R)
\end{center}
Then server collects $y_i$ and computes:
\begin{align*}
    z&=\sum\limits_{i\in n} y_i \\
    &=\sum\limits_{i\in n} \left( x_i + \sum\limits_{j\in n:i<j} s_{i,j}-\sum\limits_{j\in n:i>j} s_{i,j} \right) \\
    &=\sum\limits_{i\in n} x_i\ \rm{(mod\ R)}
\end{align*}

\begin{algorithm}[t]
	\caption{Crypto-assisted parameter selection} 
	\label{alg:ope_enhanced} 
	\begin{algorithmic}[1] 
		\REQUIRE  
		Adjacent bit vectors $\{B_1,...B_n\}$, \\
		\qquad security parameters $a,b$
		\ENSURE  
		Projection parameter $\theta$
		
		\FOR{each integer $k \in \{1,2,...,K\}$}
		\STATE /* User side. \qquad \qquad   */
		\FOR{each user $i\in \{1,2,...,n\}$}
		\STATE $\hat{d_i} \leftarrow$ $\mathtt{LocalProjection}$($B_i$, $k$) // Sec. \ref{sec:local_projection}
		\STATE $d_i \leftarrow \sum_{j=1}^n b_{i,j}$
		\STATE $noise$ $\leftarrow randint(0,a-1)$
		\STATE r $\leftarrow$ $\mathtt{PRG}$($seed$)
		\STATE $mask=\sum_{j=i+1}^{n-1}r_{i,j}-\sum_{j=1}^{i-1}r_{i,j}$
		\STATE $Enc_{T_{k,i}} \leftarrow a*|d_i-\hat{d_i}|+b+noise+mask$
		\STATE User $i$ sends $Enc_{T_{k,i}}$ to server
		\ENDFOR \\
		/* Curator side. \qquad \qquad   */
		\STATE $Enc_{T_{k}} \leftarrow \sum_{i=1}^n Enc_{T_{k,i}}$
		\STATE $\theta \leftarrow k$ when $(Enc_{T_{k}}+E_D)$ is minimum
		\ENDFOR \\
		\RETURN $\theta$ 
	\end{algorithmic} 
\end{algorithm}

Based on above two cryptographic primitives, we propose a crypto-assisted parameter selection method, as presented in Algorithm \ref{alg:ope_enhanced}.
First, we use the linear OPE scheme \cite{6253544} to encode individual utility loss, namely, $f(x)=a*|d_i-\hat{d_i}|+b+noise$. 
Here security parameters $a$ and $b$ are kept secret from the server and the noise is randomly selected from $[0, a-1]$.
Second, to hide the order of individual utility loss, we add one mask into the encoded values of the OPE scheme using SA.
For each user $i$, it and the rest other $n-1$ users agree on common seeds.
Then local users generate the random numbers $r$ with the common seeds by the pseudorandom generator (PRG) \cite{blum1984generate} and add into the individual utility loss.
Finally, the server collects all encrypted individual projection utility loss and computes the aggregated utility loss.
The added masks can be cancelled with each other after aggregation and any information about individuals cannot be leaked.
The final aggregated utility loss is still protected under OPE scheme.

\section{Local Projection Methods}
\label{sec:local_projection}
\subsection{Node-level Local Projection}
\label{sec:node-level}
Local scenarios make projection operations challenging, since no party owns the entire graph and local users cannot easily add or remove any edges. 
We propose a node-level projection method where each node is the minimal unit. 
As presented in Algorithm \ref{alg:nodeprojection}, it inputs an adjacent bit vector and projection parameter $\theta$. 
Each local user first counts the number of neighboring edges.
If node degree $d_i$ is larger than $\theta$, projected degree $\hat{d_i}$ will be directly set as $\theta$; 
otherwise, $\hat{d_i}$ remains the original value.

\begin{algorithm}[h]
	\caption{Node-level Local Projection} 
	\label{alg:nodeprojection} 
	\begin{algorithmic}[1] 
		\REQUIRE  
		Adjacent bit vector $B_i$=$\{b_{i1},...,b_{in}\}$,\\ \qquad projection parameter $\theta$
		\ENSURE  
		$\theta$-bounded node degree $\hat{d_i}$
		\STATE $d_i \leftarrow \sum_{j=1}^n b_{i,j}$
		
		\IF{$d_i > \theta$}
		\STATE $\hat{d_i} = \theta$ 
		\ELSE
		\STATE $\hat{d_i} \leftarrow d_i$
		\ENDIF
		\RETURN $\hat{d_i}$
	\end{algorithmic} 
\end{algorithm}

\begin{figure}[t]
	\centering  
	\includegraphics[width=0.8\linewidth]{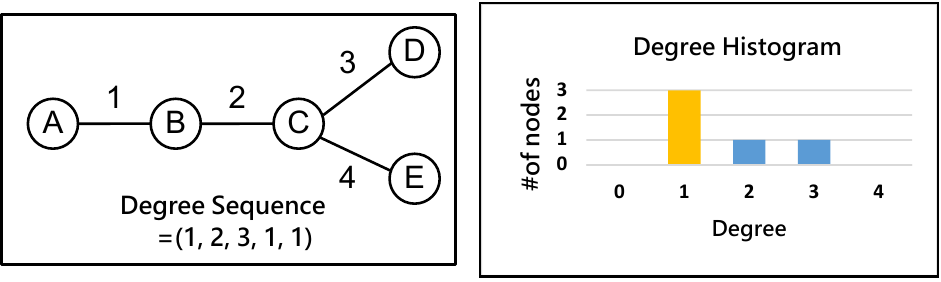}
	\caption{Example of degree histogram}  
	\label{fig:node_level}  
\end{figure}

\textbf{Limitations.} 
Although node-level projection is easy to implement, it omits correlations among neighboring users coarsely, influencing the accuracy significantly. 
For example, we have a simple graph with five nodes and some edges, as shown in Fig.  \ref{fig:node_level}. 
The original histogram can be represented as $H_1$ = (0, 3, 1, 1, 0). 
Assume that the projection parameter $\theta$ = 1, the projected degree sequence becomes $Seq_1$ = (1, 1, 1, 1, 1) and the current histogram is $H_2$= (0, 5, 0, 0, 0) after node-level projection. 
We can compute the projection loss: MSE($H_1$, $H_2$)= $\frac{6}{5}$. 
If correlations are considered, any change in mutual edges will update two neighboring adjacent bit vectors.
For example, if edge 2 and 3 are removed to bound all degrees, the degree sequence will become $Seq_2$= (1, 1, 1, 0, 1) and the degree histogram will be $H_3$= (1, 4, 0, 0, 0).
The projection loss can be computed: MSE($H_1$, $H_3$)= $\frac{4}{5}$.
Obviously, node-level method loses more edge information, which significantly affects overall utility.
What's more, the characteristic of degree distribution is destroyed by node-level projection. For instance, it is not easy to find a real-world graph that is represented by the sequence $Seq_1$. 

Generally, we assume that the number of users in a graph is $n$, projection parameter is $\theta$, and original degree histogram is $H_1$=($h_1, h_2, ..., h_n$). 
If there are $m$ nodes with degree larger than $\theta$, we can get the projected histogram $H_2$=($h_1,h_2,...,h_\theta+m,0,...,0$) using node-level projection.
On the other hand, if mutual edge information is considered during the projection, the new histogram will become $H_3$=($h_1+t_1,h_2+t_2,...,h_\theta+t_m,0,...,0$), where $t_i\in\mathbb{Z}$ ($i\in[1,m]$) is the variation of each bin in the histogram.
We refer to this method as edge-level projection method.
One mutual edge connects two nodes and there are two cases during the edge-level local projection:
(1) two node degrees are both over $\theta$. The final histogram of edge-level is same with that of node-level.
(2) one node degree is larger than $\theta$ and the other one is smaller than $\theta$. 
The change from the former one is same with the first case.
The influence from the latter can be cancelled finally.
Thus, we can easily achieve $m=t_1+t_2+...+t_m$. 
Then we can compute their projection loss, namely, $MSE(H_1, H_2$)=$\frac{m^2}{n}=\frac{1}{n}(t_1+t_2+...+t_m)^2$ and $MSE(H_1, H_3$)=$\frac{1}{n}(t_1^2+t_2^2+...+t_m^2)$. 
Since $(t_1+t_2+...+t_m)^2 \geq (t_1^2+t_2^2+...+t_m^2)$, we can get $MSE(H_1,H_2) \geq MSE(H_1,H_3)$.
Therefore, the result of node-level projection method is not desirable.

\subsection{Edge-level Local Projection}
Based on above discussions, if we consider the correlation among users, more edge information will be reserved after the projection.
However, unlike Node-CDP where the trusted server can decide the optimal strategies of removing which edges or nodes to maximize the overall utility, it is difficult for a local user to update the mutual edges. 
The key challenge is that any change in the edges may leak individual sensitive information via mutual edges. 
For example, if one node degree $d_i$ is larger than $\theta$, it will delete some edges. 
At the same time, this user $i$ will send messages to its neighboring users to update their adjacent bit vectors.
The message itself reveals that the current node degree may be larger than $\theta$. 
We design an edge-level method to protect this sensitive message.

\begin{algorithm}[t]
	\caption{Edge-level Local Projection} 
	\label{alg:edgeprojection} 
	\begin{algorithmic}[1] 
		\REQUIRE  
		Adjacent bit vector $B_i$=$\{b_{i1},...,b_{in}\}$,\\ \qquad projection parameter $\theta$, privacy budget $\varepsilon_2$
		\ENSURE  
		$\theta$-bounded node degree $\hat{d_i}$
		
		\STATE R$_i$=[0] $\times$ $\hat{d_i}$  // Record which edges will be deleted
		\STATE $d_i \leftarrow \sum_{j=1}^n b_{i,j}$
		\IF{$d_i \geq \theta$}
		\STATE Randomly select $(d_i - \theta)$ bits from R$_i$ and set '1'
        \FOR{each r$_{ij}$ $\in$ R$_i$}		
		\STATE\[
         r_{ij}^\prime=\left \{
         \begin{array}{ll}
            r_{ij}  & w.p.\ \frac{\theta}{d_i} \\
            1-r_{ij}  & w.p.\ \frac{d_i-\theta}{d_i}
         \end{array}
        \right.
       \]
		\ENDFOR
		
		\ELSE
		\FOR{each r$_{ij} \in$ R$_i$}
		\IF{$\frac{d_i-\theta}{d_i} \leq \frac{e^{\varepsilon_2}-1}{e^{\varepsilon_2}-e^{-\varepsilon_2}}$ }
		\STATE\[
         r_{ij}^\prime=\left \{
         \begin{array}{ll}
            r_{ij}  & w.p.\ 1-\frac{e^{-\varepsilon_2}(d_i-\theta)}{d_i} \\
            1-r_{ij}  & w.p.\ \frac{e^{-\varepsilon_2}(d_i-\theta)}{d_i}
         \end{array}
        \right.
       \]
        \ELSE
        \STATE\[
         r_{ij}^\prime=\left \{
         \begin{array}{ll}
            r_{ij}  & w.p.\ \frac{e^{\varepsilon_2}\theta}{d_i} \\
            1-r_{ij}  & w.p.\ \frac{d_i-e^{\varepsilon_2}\theta}{d_i}
         \end{array}
        \right.
       \]
		\ENDIF
		\ENDFOR
		\ENDIF
        \FOR{each r$_{ij}$ $\in$ R$_i$}	
        \IF{$r_{ij} =1$}
        \STATE $b_{ij}=0$ and $b_{ji}=0$
        \ENDIF
        \ENDFOR
		\RETURN $\hat{d_i}$
	\end{algorithmic} 
\end{algorithm}

\textbf{Security Assumptions.}
We assume that 1) the communication between neighboring users is perfectly anonymous, that's to say, the third party (e.g., server or third user) cannot know the communication exists or not; 
2) the user does not reveal sensitive neighboring information to other users, for example, B will not tell C that A is one of its friends or not.
Based on above assumptions, one edge is only visible to two neighboring users and other edges are in a data-invisible way.
Thus, the communication message is just one bit and the sensitivity becomes $O(1)$.

\begin{table}[t]
	\caption{Randomized projection vector}
	\label{tab:projection_probability}
	\centering
	\setlength{\tabcolsep}{5mm}{
		\begin{tabular}{lrr}
			\hline
			Pr & 0 &1 \\
			\hline
			$\textless \theta$ & 1-x & x\\
			$\geq \theta$ & 1-p & p \\
		    \hline
	\end{tabular}}
\end{table}

\textbf{Algorithm.} 
We propose the edge-level projection method to improve node-level method and the edge is the minimal unit during the projection, as shown in Algorithm \ref{alg:edgeprojection}. 
Privacy leakage may occur when the local projection is performed since the sensitive messages are sent to neighboring users.
We represent this message as an operation vector $R_i=\{r_{i1},..., r_{id_i}\}$, and the size of $R_i$ is $d_i$. 
If $r_{ij}=1$, the corresponding edges in two neighbor lists will be removed; otherwise, they remain the same. 
We carefully perturb each bit of the operation vector to make two cases indistinguishable: node degree $d_i$ is larger than $\theta$ or $d_i$ is smaller than $\theta$.
Ideally, we want to flip each bit of the projection bit vector with probability in Table~\ref{tab:projection_probability}, where $p=\frac{d_i-\theta}{d_i}$ and $x=0$. Obviously, when $x=0$, our randomized mechanism cannot satisfy the Node-LDP.
To satisfy the Node-LDP, we have the following inequation:

\begin{equation}
\label{equation:edge_level}
    \left \{
        \begin{array}{ll}
            e^{-\varepsilon_2} \leq \frac{x}{p} \leq e^{\varepsilon_2}  \\
            e^{-\varepsilon_2} \leq \frac{1-x}{1-p} \leq e^{\varepsilon_2} 
         \end{array}
        \right. \nonumber
\end{equation}
Then, we can bound the scope of $x$ as follows:
\begin{equation}
    \left \{
        \begin{array}{cc}
            pe^{-\varepsilon_2} \leq x \leq pe^{\varepsilon_2}  \\
            (p-1)e^{\varepsilon_2}+1 \leq x \leq (p-1)e^{-\varepsilon_2}+1 
         \end{array}
        \right.\nonumber
\end{equation}
When $d_i \textless \theta$, we want to preserve more edges during projection, that is to say, the number of `1' in projection bit vector is as small as possible. Thus we have
\begin{equation}
    x=\left \{
        \begin{array}{lc}
            pe^{-\varepsilon_2}, & pe^{-\varepsilon_2} \geq (p-1)e^{\varepsilon_2}+1  \\
            (p-1)e^{\varepsilon_2}+1, & pe^{-\varepsilon_2} \textless (p-1)e^{\varepsilon_2}+1
         \end{array}
        \right.\nonumber
\end{equation}

After randomizing the bits of the projection bit vector, each user updates the adjacent bit vector according to randomized bit vector (Line 19). 
Then, local users count the number of edges and obtain the bounded degree $\hat{d_i}$.

\section{Analysis and Discussions}
\label{sec:analysis discussion}
\textbf{Privacy Budget Allocation.}
As shown in Algorithm \ref{alg:degree_distribution}, there are three kinds of privacy budgets.
Our goal is to find the optimal privacy allocation scheme with the best utility.
Without loss of generality, we assume that the overall privacy budget is $\varepsilon$, $\varepsilon_3=\alpha\varepsilon$, and $\varepsilon_1+\varepsilon_2=(1-\alpha)\varepsilon$.
For inner privacy budget allocation of local graph projection, we distribute the same privacy budget for the projection parameter selection and executing the local graph projection, namely, $\varepsilon_1=\varepsilon_2$.
We find the optimal $\alpha$ with the least utility loss by conducting many experiments for different cases, as shown in Table \ref{tab:privacy allocation}.
And we use the optimal $\alpha$ for each case in the next experiments.

\textbf{Selection of Parameter $K$.}
In Algorithm \ref{alg:ldp_enhanced} and Algorithm~\ref{alg:ope_enhanced}, the parameter $K$, namely, the size of the candidate pool, plays a significant role in the tradeoff between utility and privacy.
When the size $K$ is larger, more noise will be added by the pureLDP parameter selection and time overhead becomes higher.
Similarly, the running time of crypto-assisted selection method will be higher.
But if the $K$ becomes smaller, the optimal projection parameter $\alpha$ is not covered possibly.
We conduct extensive experiments and find the optimal parameter $\alpha$ for each case, as shown in Table \ref{tab:hyper-parameter K}.
In our paper, we use $K=50$ that is ample to cover the optimal parameter $\alpha$ of different cases.

\begin{table}[t]
	\caption{optimal privacy allocation scheme $\alpha$}
	\centering
	\label{tab:privacy allocation}
	\setlength{\tabcolsep}{2mm}{
		\begin{tabular}{lcccc}
			\hline
			$\varepsilon$ & Ca-HepPh & Cit-HepPh & Twitter & Com-DBLP \\\hline
			0.5 & 0.895 & 0.927 &  0.945 & 0.945  \\
			1   & 0.944 & 0.937 &  0.949 & 0.947  \\
			1.5 & 0.901 & 0.940 &  0.944 & 0.948  \\
			2   & 0.948 & 0.946 &  0.947 & 0.937  \\
			2.5 & 0.944 & 0.922 &  0.948 & 0.943  \\
			3   & 0.944 & 0.948 &  0.941 & 0.940  \\\hline
	\end{tabular}}
\end{table}

\begin{table}[t]
	\caption{optimal parameter $\theta$}
	\centering
	\label{tab:hyper-parameter K}
	\setlength{\tabcolsep}{2mm}{
		\begin{tabular}{lcccc}
			\hline
			$\varepsilon$ & Ca-HepPh & Cit-HepPh & Twitter & Com-DBLP \\\hline
			0.5 & 3 & 4&  18& 13  \\
			1   & 9 & 7&  31& 17  \\
			1.5 & 15& 10& 41& 20 \\
 			2   & 19& 12& 43& 23 \\
			2.5 & 24& 15& 45& 25 \\
			3   & 26& 18& 48& 27 \\\hline
	\end{tabular}}
\end{table}

\textbf{Time Complexity.} 
As shown in Table \ref{tab:running time}, we conclude the running time complexity of different combinations theoretically, $|V|$ and $|E|$ represent the number of nodes and edges respectively.
Node-level local projection method transforms each node degree into $\theta$-bounded degree directly, which takes time $O(|V|)$.
In contrast, edge-level local projection method needs to traverse each edge for each node, resulting an $O(|V|.|E|)$ running time.
PureLDP parameter selection method selects the optimal parameter $\theta$ from $K$ candidates and for each candidate $k$, each user has to compute the projection loss, which takes time at most $O(K.|V|)$.
By comparison, for each candidate parameter $k$ of crypto-assisted selection method, each user has to communicate with the other $(|V|-1)$ users to determine the seed, resulting an $O(K.|V|^2)$ running time overhead.

\begin{table}[htb]
	\caption{running time complexity}
	\centering
	\label{tab:running time}
	\setlength{\tabcolsep}{1mm}{
		\begin{tabular}{lcc}
			\hline
			 & pureLDP & crypto-assisted \\\hline
			Node-level & \small$O(|V|+K.|V|)$ & \small$O(|V|+K.|V|^2)$  \\
			Edge-level & \small$O(|V|.|E|+K.|V|)$ & \small$O(|V|.|E|+K.|V|^2)$  \\
			\hline
	\end{tabular}}
\end{table}

\begin{figure*}[t]
	\centering
	\subfigure{
		\begin{minipage}[t]{\linewidth}
			\centering
			\includegraphics[width=0.7\linewidth]{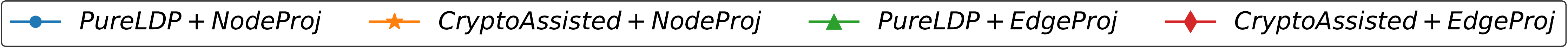}
		\end{minipage}
	}%
	\qquad
	\subfigure[Ca-HepPh]{
		\begin{minipage}[t]{0.238\linewidth}
			\centering
			\includegraphics[width=\linewidth]{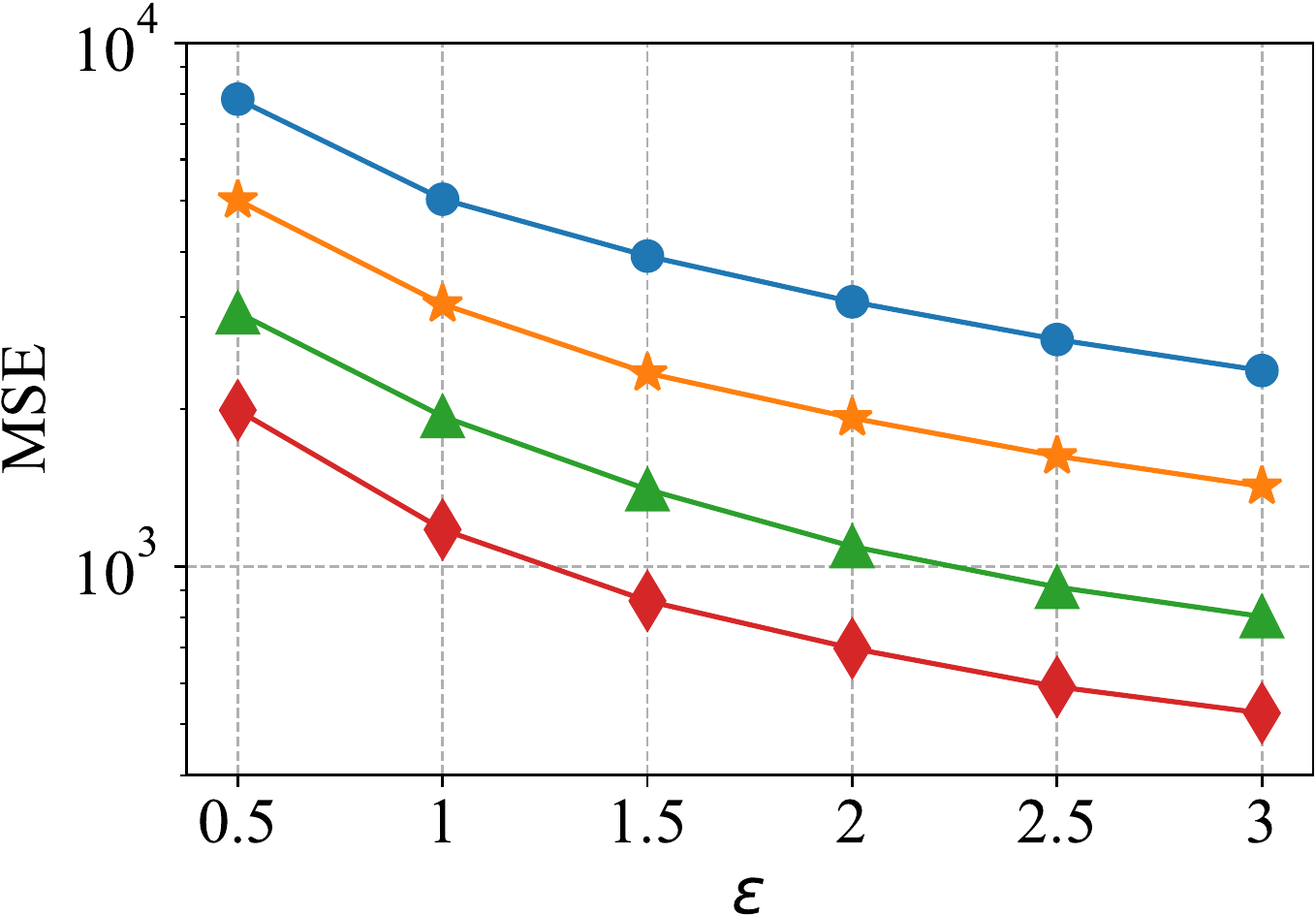}
			\label{fig:mse_epsilmse_epsilon_caheph}
		\end{minipage}
	}%
	\subfigure[Cit-HepPh]{
		\begin{minipage}[t]{0.238\linewidth}
			\centering
			\includegraphics[width=\linewidth]{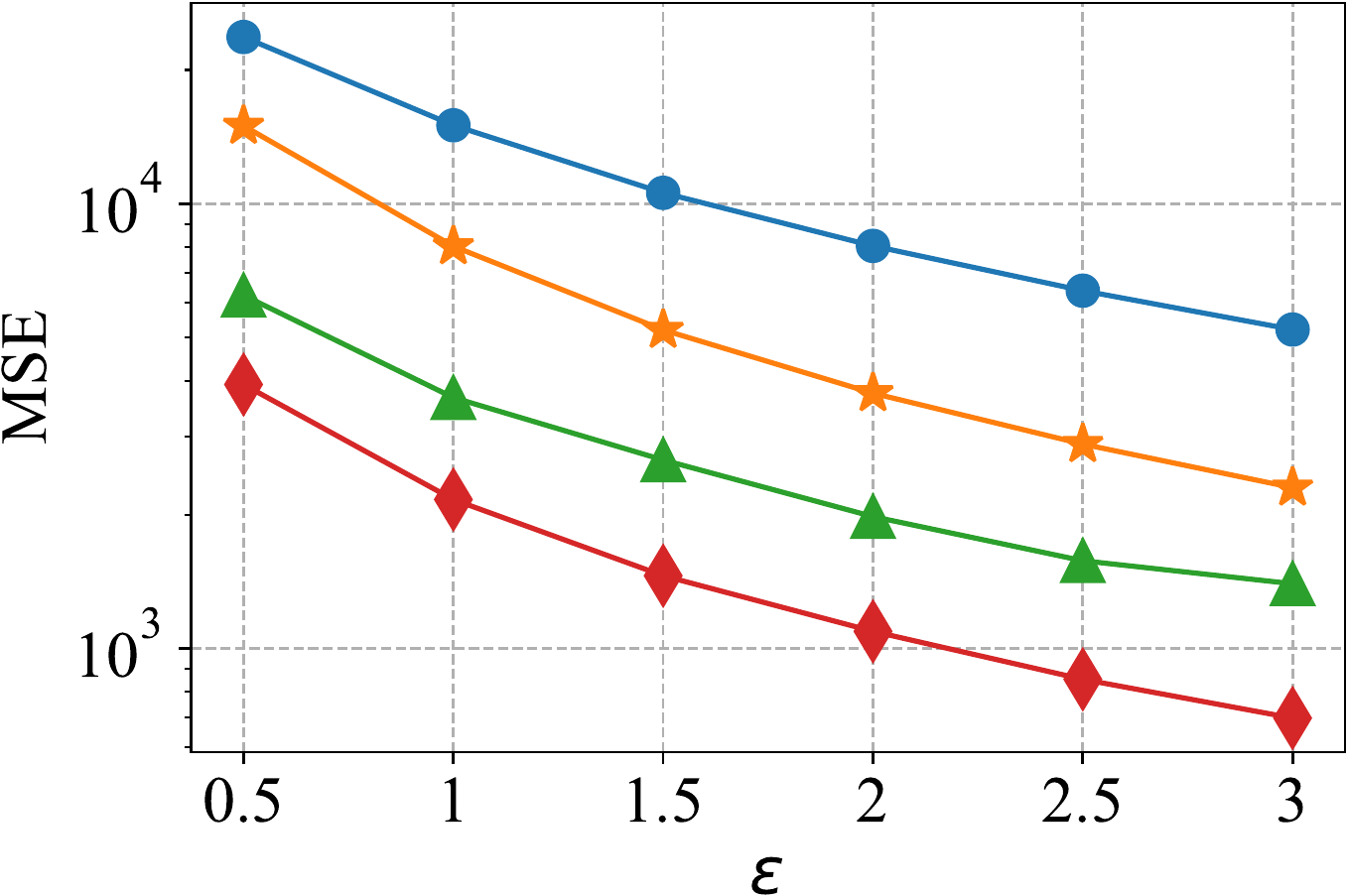}
			\label{fig:mse_epsilon_citheph}
		\end{minipage}%
	}
	\subfigure[Twitter]{
		\begin{minipage}[t]{0.238\linewidth}
			\centering
			\includegraphics[width=\linewidth]{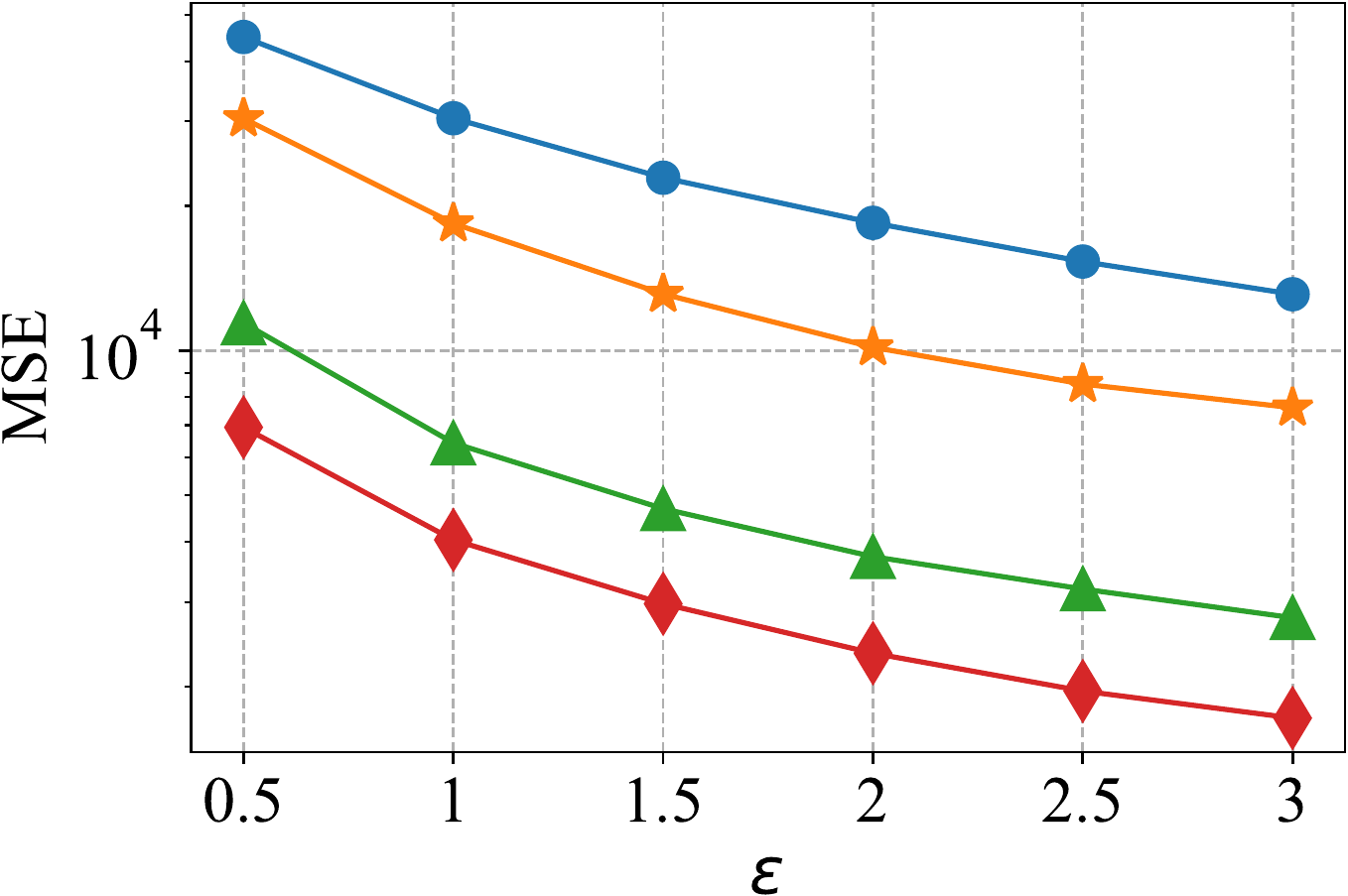}
			\label{fig:mse_epsilon_twitter}
		\end{minipage}%
	}
	\subfigure[Com-DBLP]{
		\begin{minipage}[t]{0.238\linewidth}
			\centering
			\includegraphics[width=\linewidth]{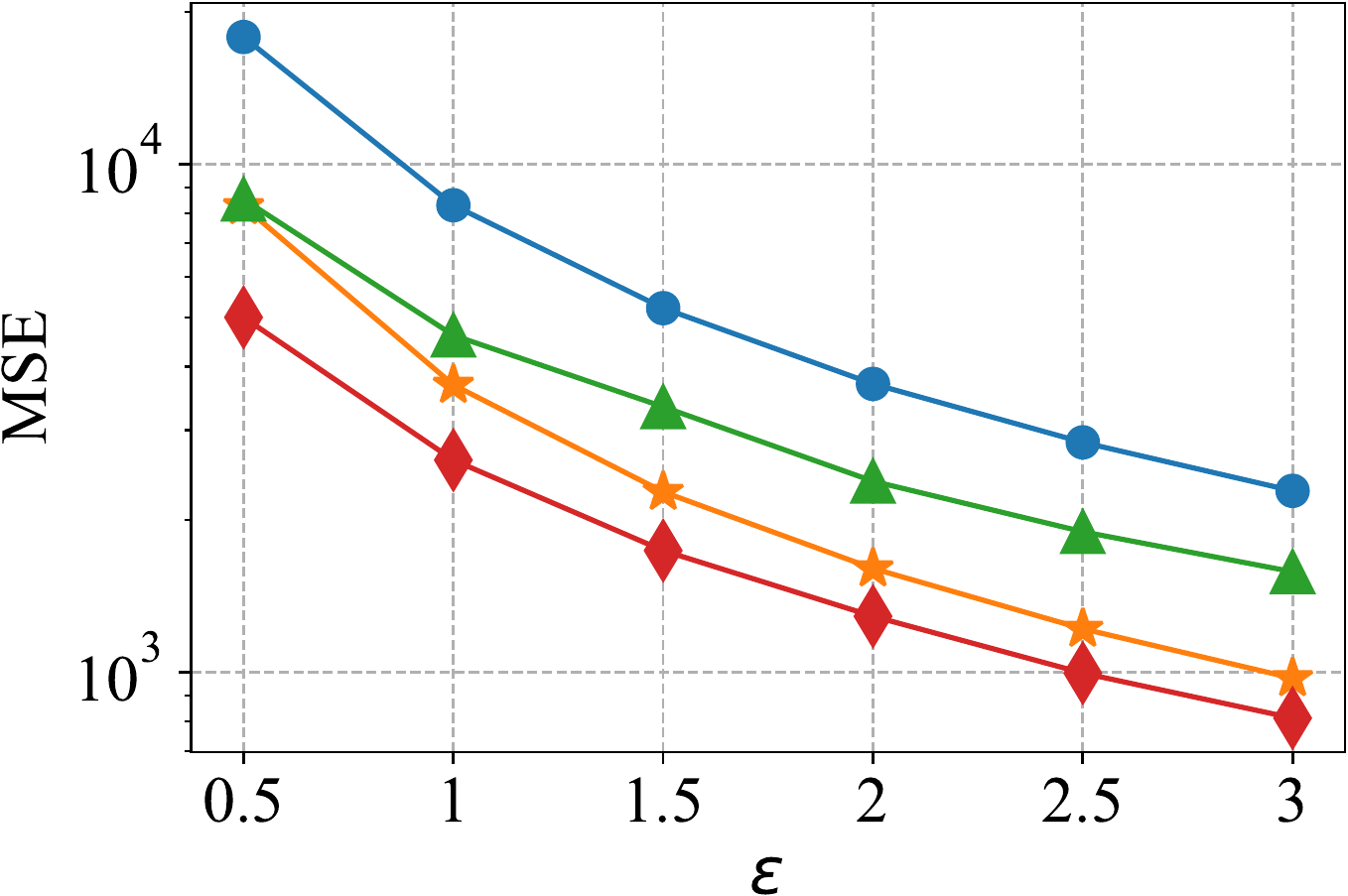}
			\label{fig:mse_epsilon_dblp}
		\end{minipage}%
	}
	\qquad
	\subfigure[Ca-HepPh]{
		\begin{minipage}[t]{0.238\linewidth}
			\centering
			\includegraphics[width=\linewidth]{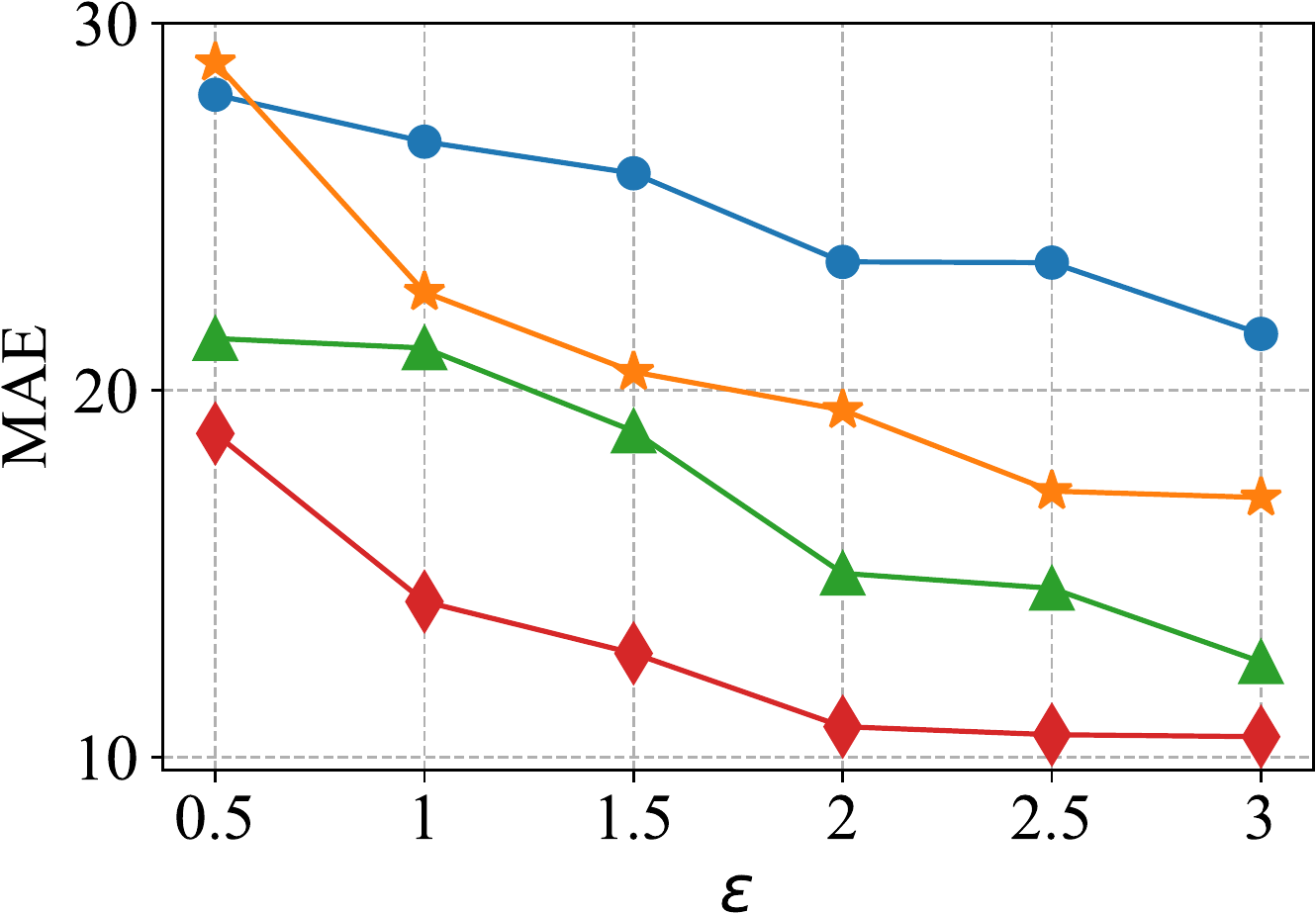}
			\label{fig:mae_epsilon_caheph}
		\end{minipage}
	}%
	\subfigure[Cit-HepPh]{
		\begin{minipage}[t]{0.238\linewidth}
			\centering
			\includegraphics[width=\linewidth]{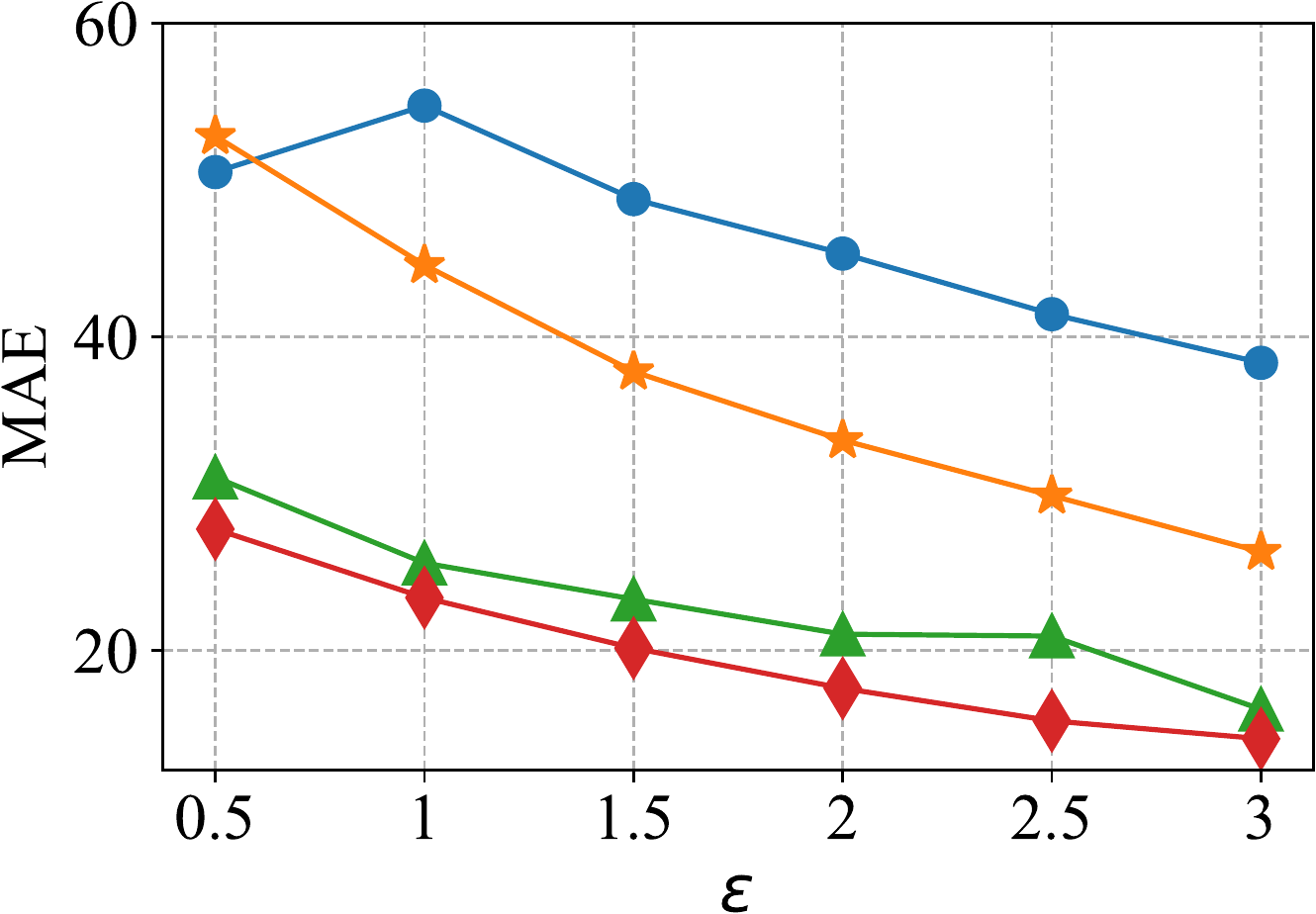}
			\label{fig:mae_epsilon_citheph}
		\end{minipage}%
	}
	\subfigure[Twitter]{
		\begin{minipage}[t]{0.238\linewidth}
			\centering
			\includegraphics[width=\linewidth]{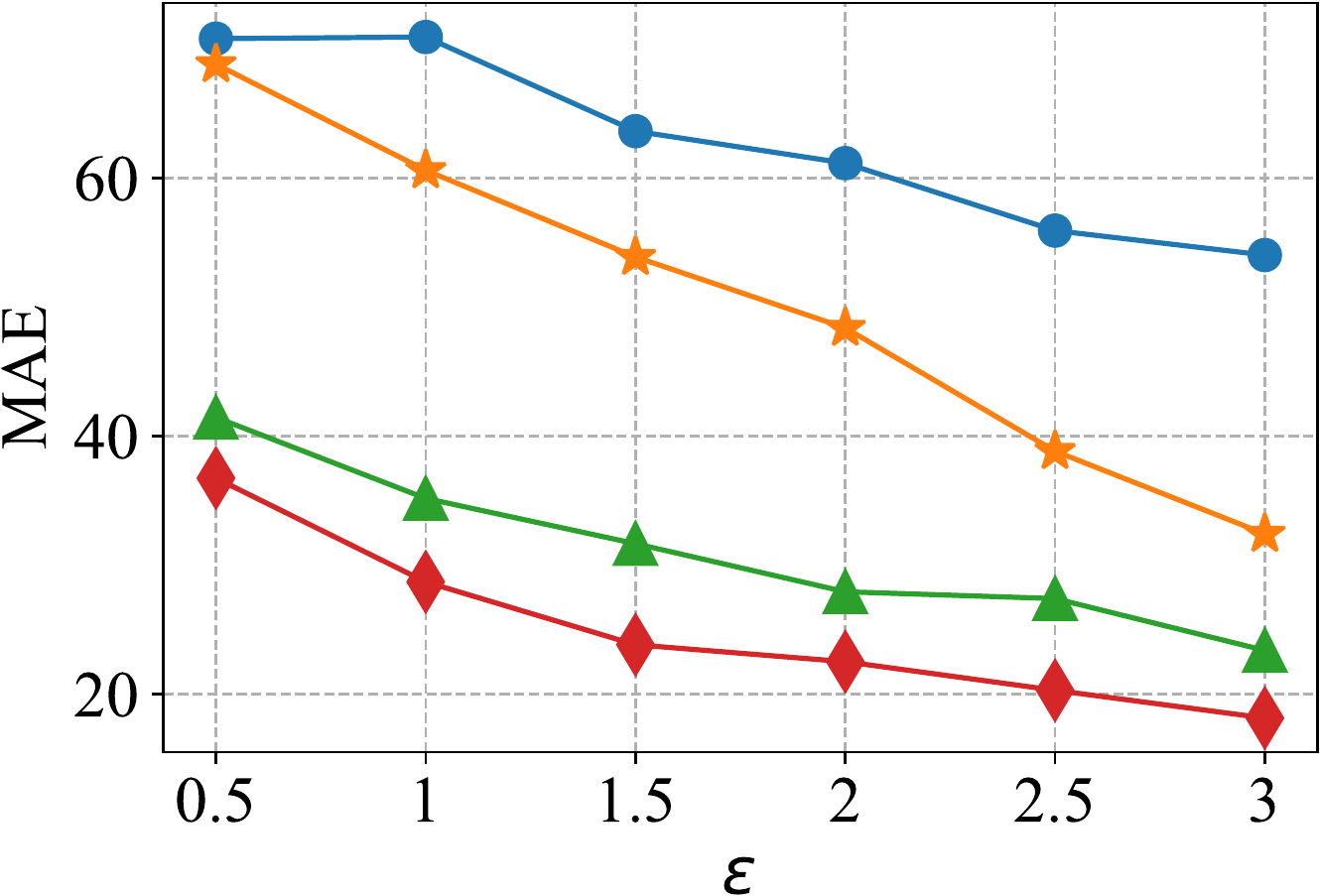}
			\label{fig:mae_epsilon_twitter}
		\end{minipage}%
	}
	\subfigure[Com-DBLP]{
		\begin{minipage}[t]{0.238\linewidth}
			\centering
			\includegraphics[width=\linewidth]{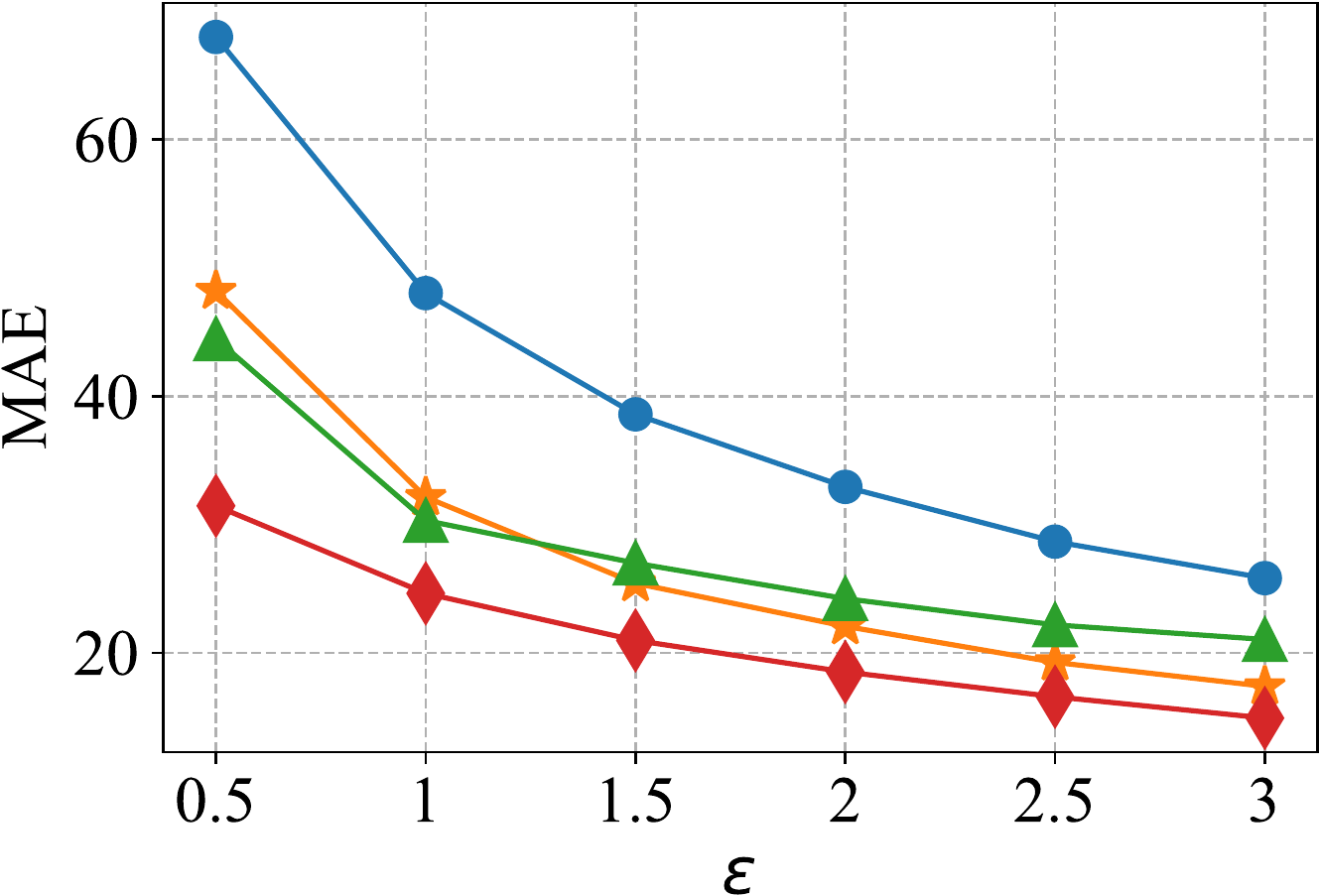}
			\label{fig:mae_epsilon_dblp}
		\end{minipage}%
	}
\centering
\caption{The MSE and MAE of algorithms on different graphs}
\label{fig:mse_mae}
\end{figure*}
\textbf{Security Analysis.} Publishing the degree distribution in Algorithm \ref{alg:degree_distribution} is under the following privacy guarantee. 
\begin{lemma}
	\label{lemma:privacy}
	Publishing the degree distribution satisfies $(\varepsilon_1/K+\varepsilon_2+\varepsilon_3)$-Node-LDP.
\end{lemma}

\emph{Proof of Lemma \ref{lemma:privacy}:} In Algorithm \ref{alg:degree_distribution}, SelectParameter(.) (Line 1) uses the Laplace with privacy budget $\varepsilon_1/K$, $K$ is the number of candidate parameters. 
Executing the local projection (Line 3) uses our proposed mechanism and satisfies Node-LDP for $\varepsilon_2$. 
And publishing the distribution with Laplace Mechanism using $\varepsilon_3$.
According to the post-processing theorem and composition property, Algorithm \ref{alg:degree_distribution} satisfies $(\varepsilon_1/K+\varepsilon_2+\varepsilon_3)$-Node-LDP.

\section{Experimental Evaluation}
\label{sec:experiment}
In this section, we would like to answer the following questions:
\begin{itemize}
  \item What is the tradeoff between utility and privacy of our proposed methods?
  \item What are results of different privacy budget allocation schemes?
  \item How much time do our proposed algorithms take?
\end{itemize}

\subsection{ Datasets and Setting}
\begin{table}[b]
	\caption{details of graph datasets}
	\centering
	\label{tab:datasets}
	\setlength{\tabcolsep}{2mm}{
		\begin{tabular}{lrrr}
			\hline
			Graph & $|V|$  & $ |E|$  & $ |E|$ $^\prime$ \\\hline
			Ca-HepPh & 12,008 & 118,521 & 474,020  \\
			Cit-HepPh  & 34,546 & 421,578  & 843,156\\
			Twitter & 81,306 & 1,768,149 & 3,536,298  \\
			Com-DBLP & 317,080 & 1,049,866 & 2,099,732  \\\hline
	\end{tabular}}
\end{table}

Our experiments run in python on a server with Intel Core i9-10920X CPU, 256GB RAM running Ubuntu 18.04 LTS. We use four real-world graph datasets from SNAP \cite{snapnets}, which are also used in \cite{day2016publishing,LF-GDPR}. 
And we preprocess all graph datasets to be undirected and symmetric graphs. 
Table \ref{tab:datasets} presents more details about every graph dataset, including the number of nodes $|V|$, the number of edges $|E|$, and the number of edges after preprocessing $|E^\prime|$ after preprocessing. 
In all experiments, we vary the privacy budget $\varepsilon$ from 0.5 to 3. By default, we set hyper-parameter $K$=50 as we discussed above. 
All of our experimental results are the average values computed from 20 runs. 
We use `PureLDP', CryptoAssisted', `NodeProj' and `EdgeProj' to represent pureLDP parameter selection, crypto-assisted parameter selection, node-level local graph projection and edge-level local graph projection respectively.
Thus we have four different combinations to publish the degree distribution.

\begin{figure*}[t]
	\centering
	\subfigure[Ca-HepPh]{
		\begin{minipage}[t]{0.238\linewidth}
			\centering
			\includegraphics[width=\linewidth]{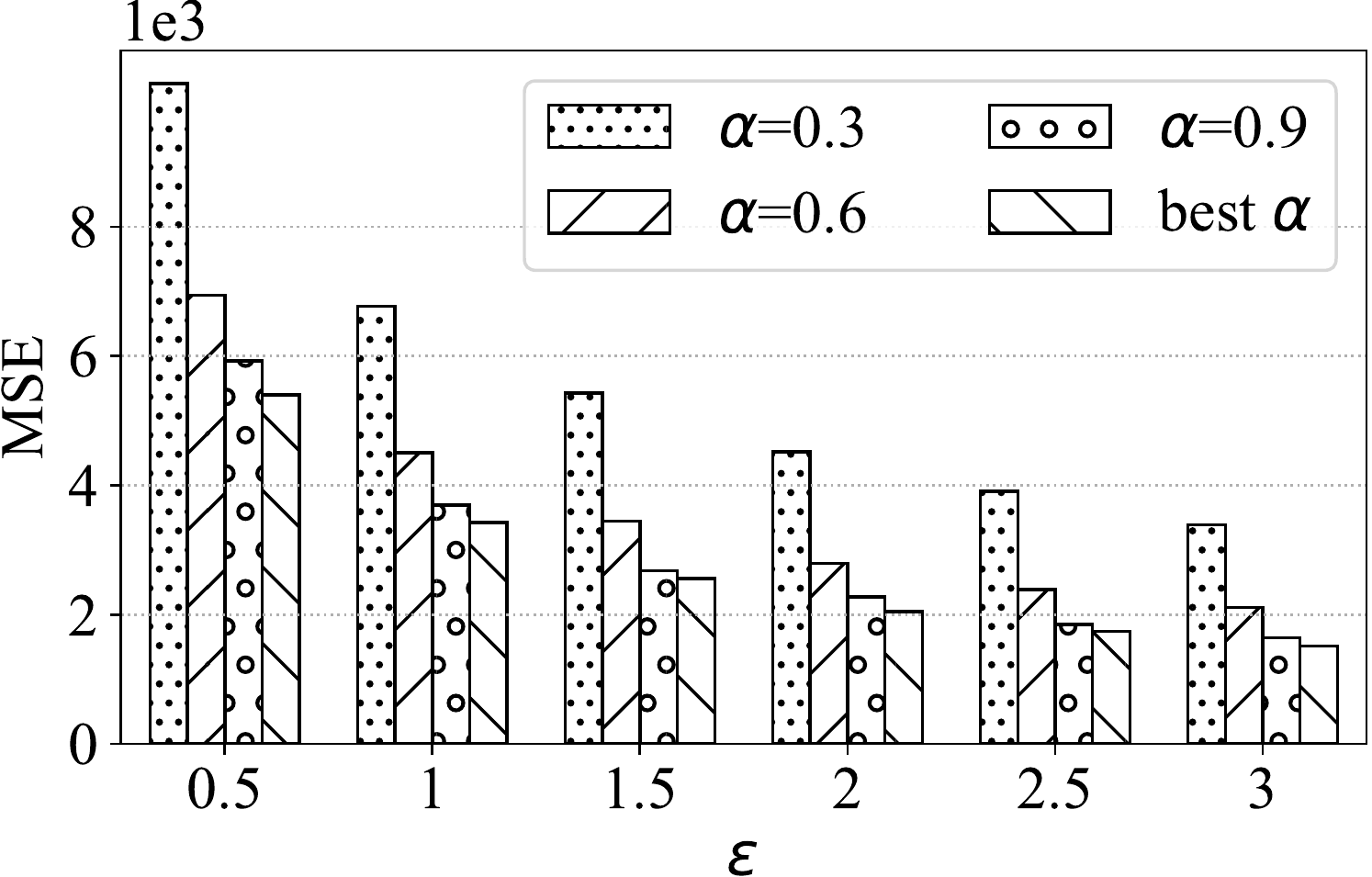}
		\end{minipage}
	}%
	\subfigure[Cit-HepPh]{
		\begin{minipage}[t]{0.238\linewidth}
			\centering
			\includegraphics[width=\linewidth]{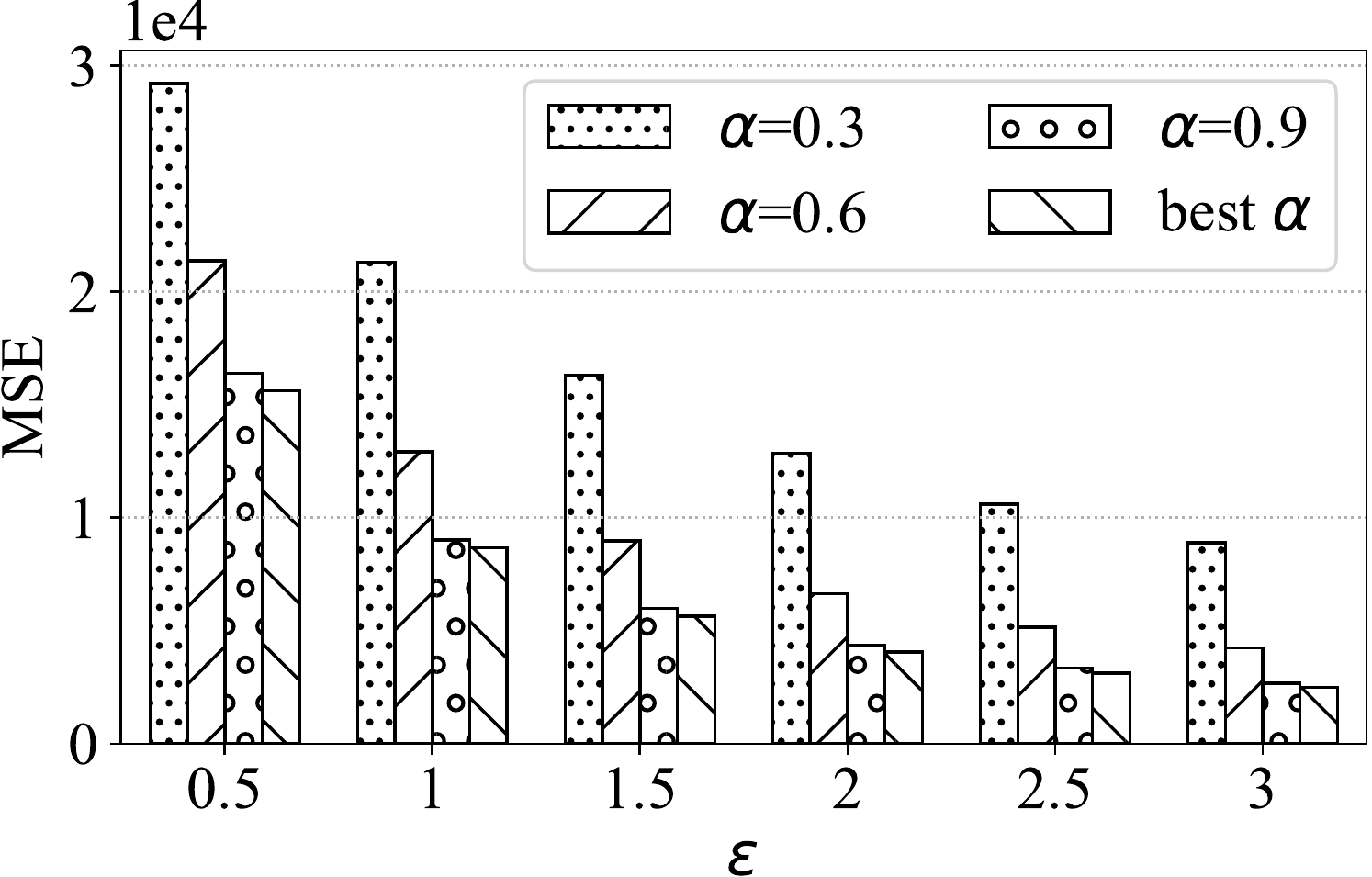}
		\end{minipage}%
	}
	\subfigure[Twitter]{
		\begin{minipage}[t]{0.238\linewidth}
			\centering
			\includegraphics[width=\linewidth]{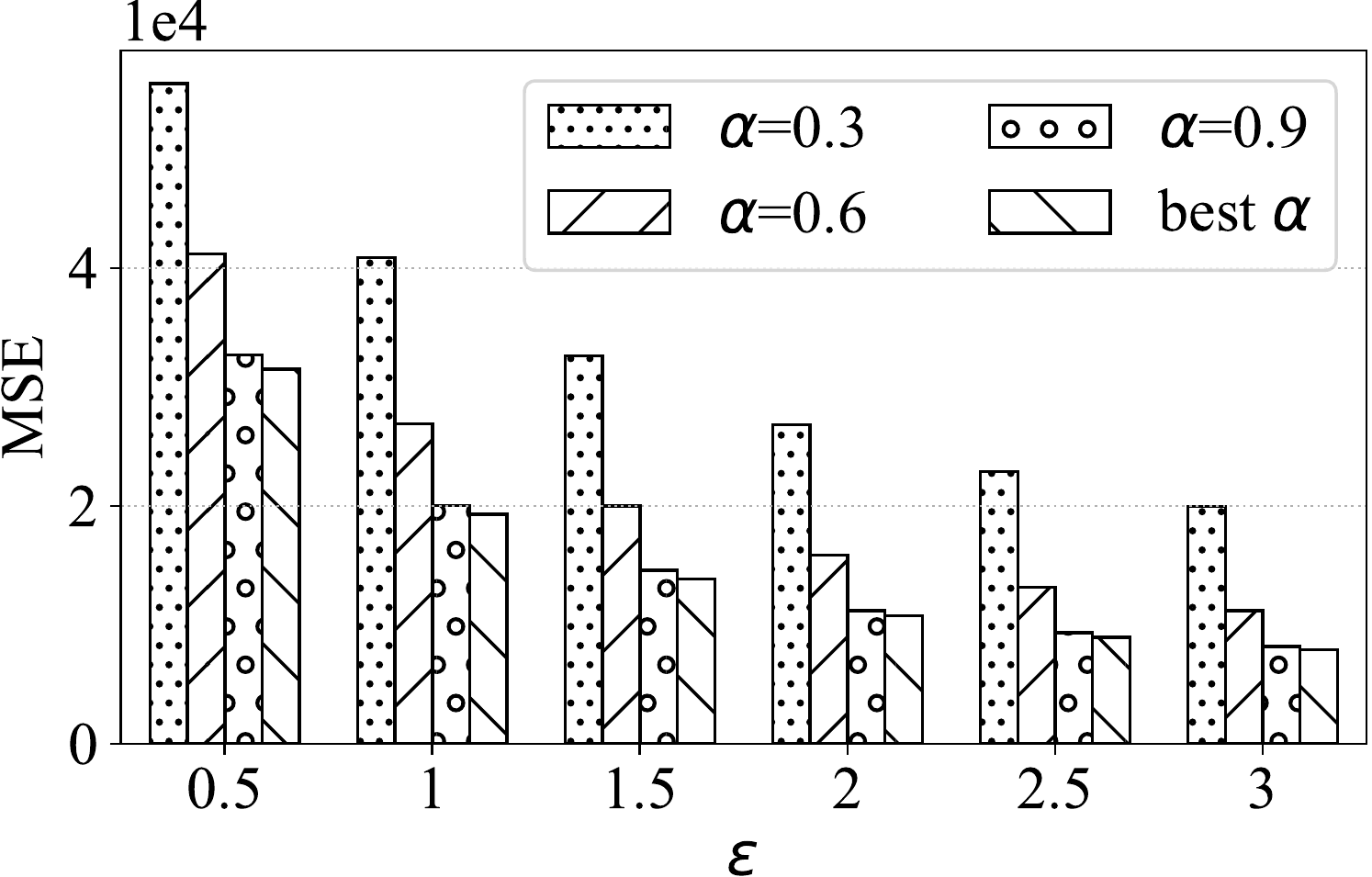}
		\end{minipage}%
	}
	\subfigure[Com-DBLP]{
		\begin{minipage}[t]{0.238\linewidth}
			\centering
			\includegraphics[width=\linewidth]{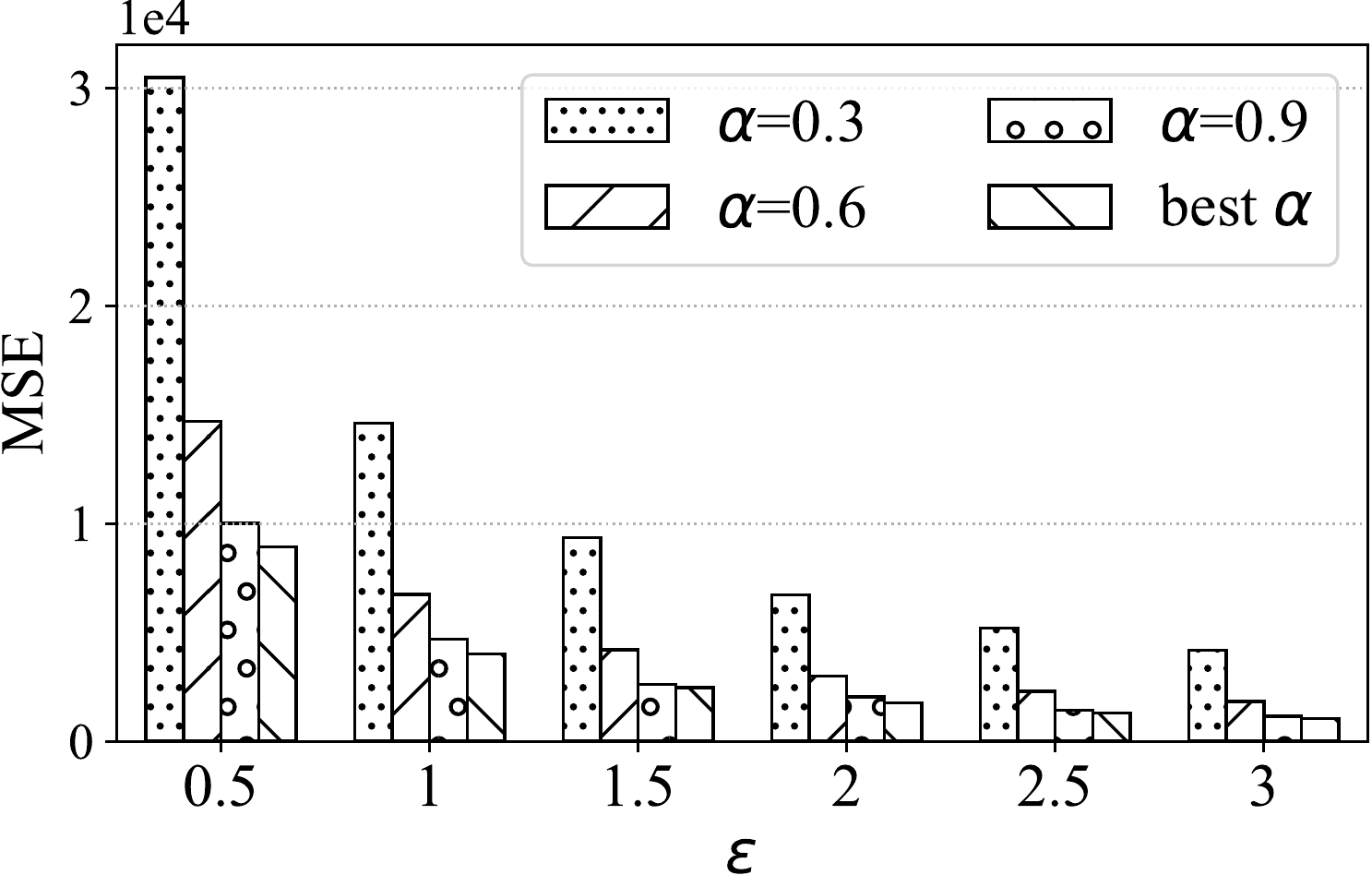}
		\end{minipage}%
	}
\centering
\caption{The MSE on different graphs, varying $\alpha$}
\label{fig:alloaction}
\end{figure*}

\begin{figure*}[t]
	\centering
		\subfigure{
		\begin{minipage}[t]{\linewidth}
			\centering
			\includegraphics[width=0.7\linewidth]{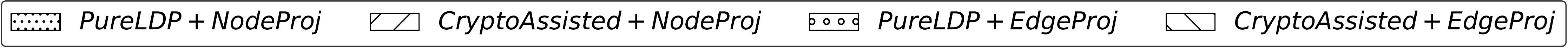}
		\end{minipage}
	}%
	\qquad
	\subfigure[Ca-HepPh]{
		\begin{minipage}[t]{0.238\linewidth}
			\centering
			\includegraphics[width=\linewidth]{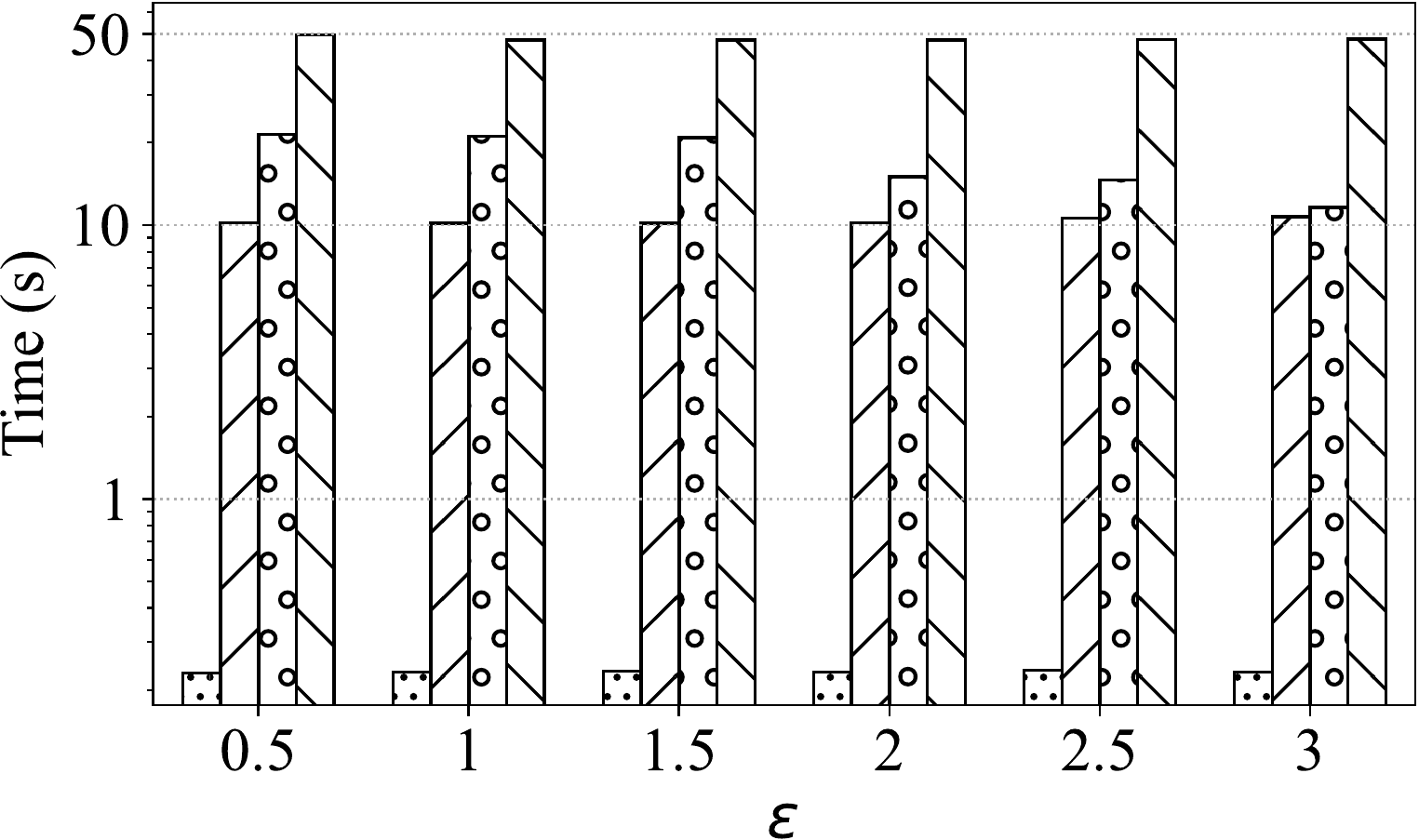}
			\label{fig:time_epsilon_caheph}
		\end{minipage}
	}%
	\subfigure[Cit-HepPh]{
		\begin{minipage}[t]{0.238\linewidth}
			\centering
			\includegraphics[width=\linewidth]{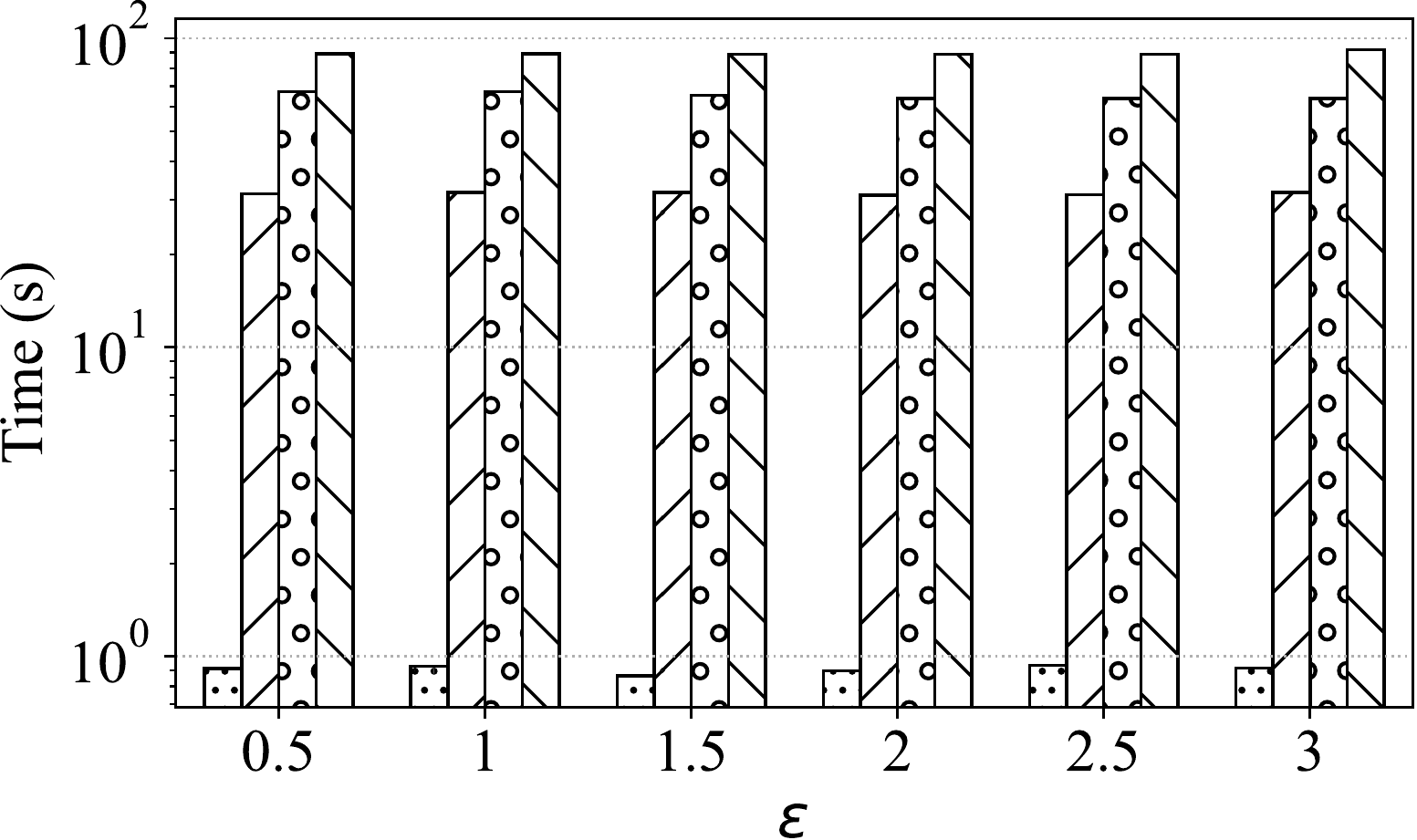}
			\label{fig:time_epsilon_citheph}
		\end{minipage}%
	}
	\subfigure[Twitter]{
		\begin{minipage}[t]{0.238\linewidth}
			\centering
			\includegraphics[width=\linewidth]{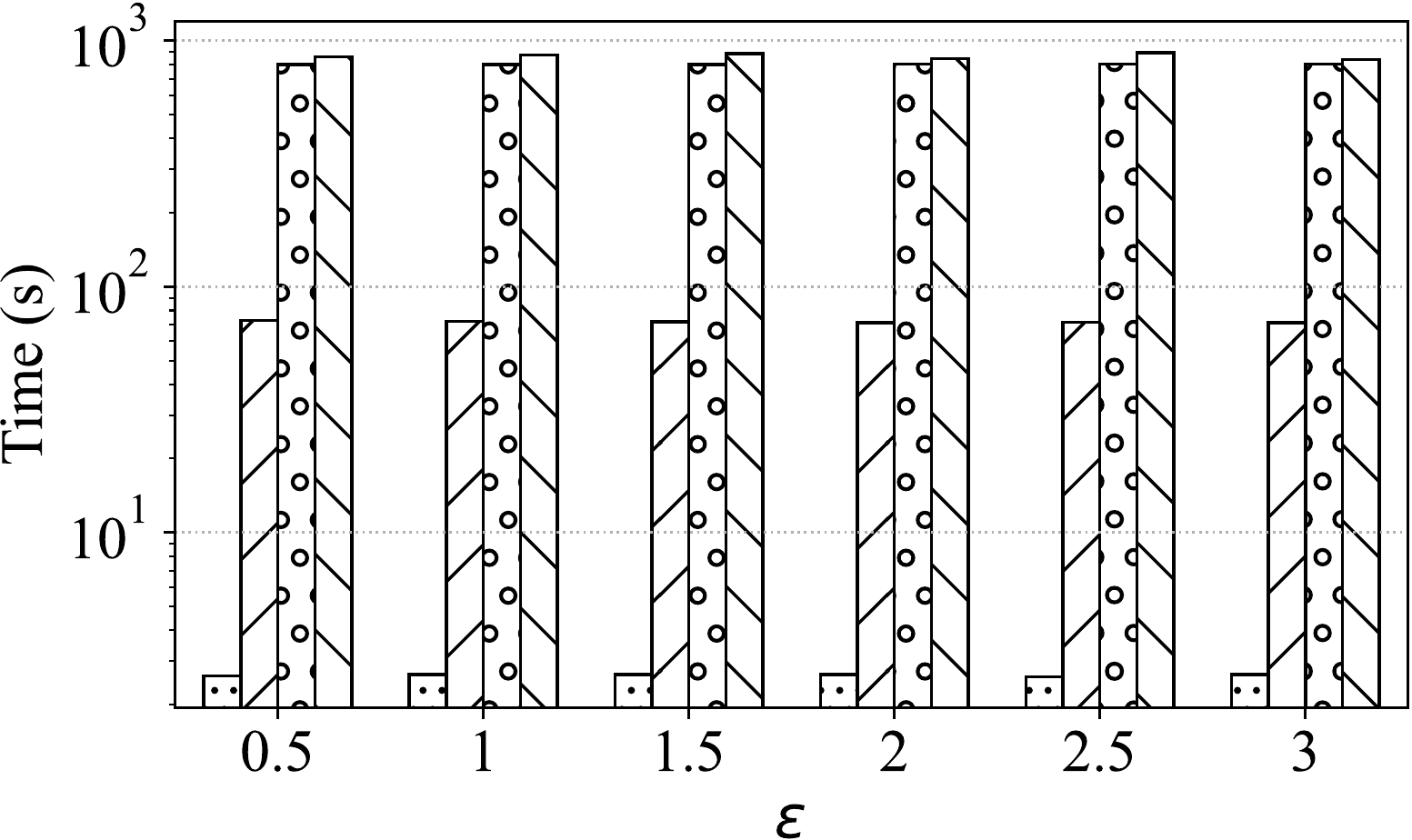}
			\label{fig:time_epsilon_twitter}
		\end{minipage}%
	}
	\subfigure[Com-DBLP]{
		\begin{minipage}[t]{0.238\linewidth}
			\centering
			\includegraphics[width=\linewidth]{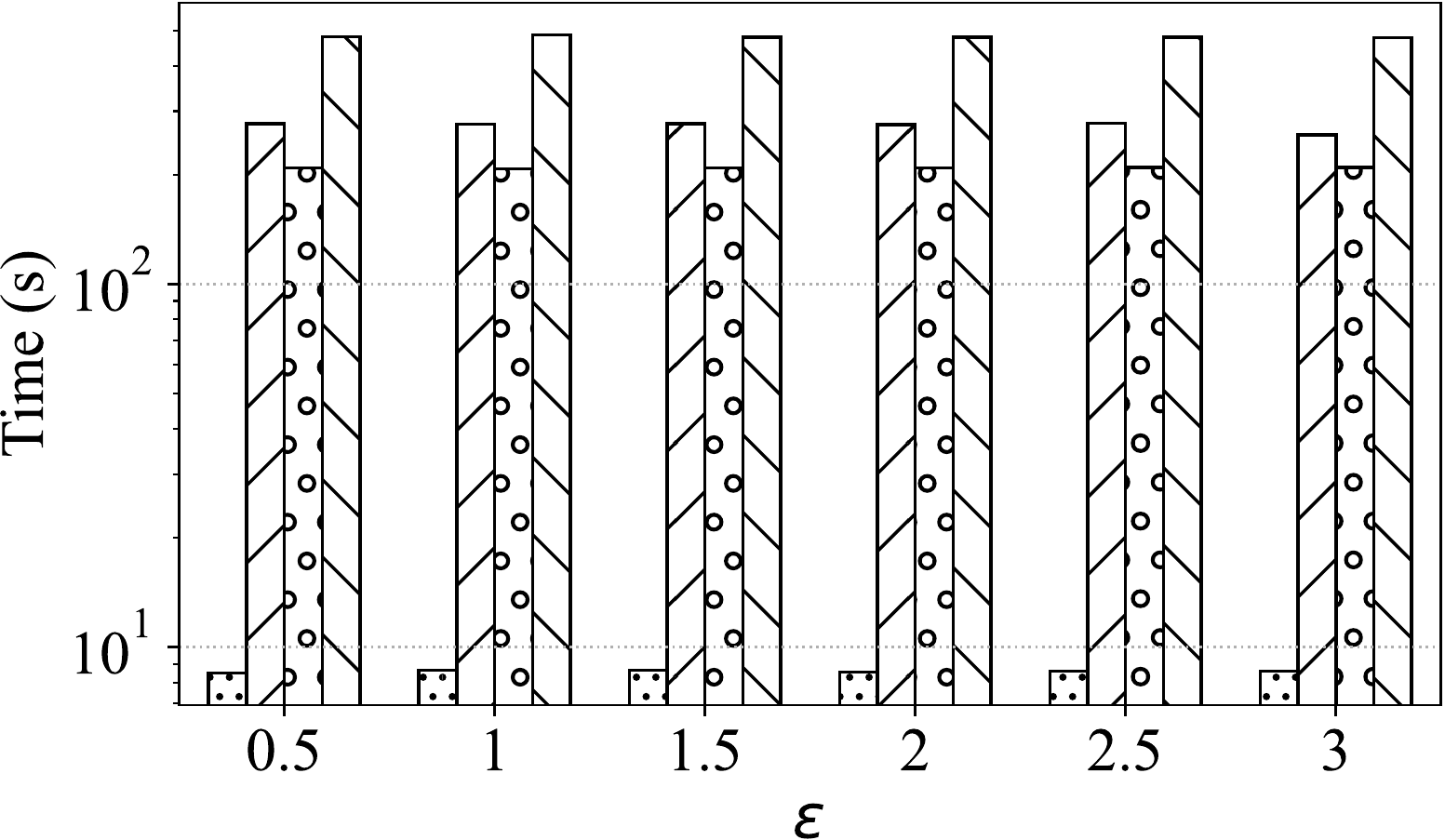}
			\label{fig:time_epsilon_dblp}
		\end{minipage}%
	}
\centering
\caption{The runtime on different graphs}
\label{fig:runtime}
\end{figure*}

\subsection{Relation between $\varepsilon$ and MSE, MAE}
As shown in Fig. \ref{fig:mse_mae}, the utility of each combination method becomes better as the privacy budget $\varepsilon$ increases.
We can find that `CryptoAssisted+EdgeProj' method always performs the best in most cases, while the results of `PureLDP+NodeProj' method are always the worst.
To be specific, the MSE of `CryptoAssisted+EdgeProj' method is less than that of `PureLDP+NodeProj' by up to 87.2\% on Twitter when $\varepsilon=2.5$.
The MAE of `CryptoAssisted+NodeProj' method is larger than that of `CryptoAssisted+EdgeProj' method by up to 66.4\% in Twitter when $\varepsilon=3$.
The reason that `CryptoAssisted+EdgeProj' method sometimes performs not the best in terms of MAE when $\varepsilon=0.5$ is because our utility loss function uses the MSE as the evaluation metric, which makes a little influence on results of MAE, particularly when $\varepsilon$ is very small.
The results of pureLDP parameter projection are always worse than that of crypto-assisted parameter projection since the latter protects individual utility loss while preserving the order information of the aggregated utility loss accurately.
Also, due to more information is preserved, edge-level local projection method performs much better than node-level local projection method.
Overall, our proposed `CryptoAssisted+EdgeProj' method improves our baseline `PureLDP+NodeProj' approach for publishing the degree distribution under Node-LDP.

\subsection{Impact of privacy budget allocation}
To further estimate the influence of the privacy allocation scheme on the overall utility, we compare the best $\alpha$ with other three constant $\alpha$, including 0.3, 0.6, and 0.9. 
We present the MSE results of different $\alpha$ on different graph datasets in Fig. \ref{fig:alloaction}. 
We can observe that the best $\alpha$ owns the lowest MSE against the other allocation schemes in most cases.
On the other hand, with the increase of the overall privacy budget $\varepsilon$, the MSE value is decreasing. 
Thus most of privacy budget can be allocated to the final publishing the degree distribution, which is roughly consistent with our best $\alpha$ in Table \ref{tab:privacy allocation}, namely, $\varepsilon_3$ for publishing degree distribution is approximately equal to the overall privacy budget $\varepsilon$.

\subsection{Analysis on running time}
Finally, we compare the running time overhead of our proposed methods, as shown in Fig. \ref{fig:runtime}.
We can see that the running time of `CryptoAssisted+EdgeProj' method is much larger than that of `PureLDP+NodeProj' method.
This is mainly because edge-level projection method needs to traverse each edge of every node and crypto-assisted parameter selection method has $n$ users to communicate in pairs, which is in line with our theoretical analysis in Section \ref{sec:analysis discussion}.
The difference between  `CryptoAssisted+EdgeProj' method and `PureLDP+NodeProj' method is larger on Twitter.
This is because Twitter has more edges than other graphs, as described in Table \ref{tab:datasets}, which results in higher computation and communication overhead.

\section{Related Work}
\label{sec:related work}
There are many existing works related to Node-CDP and Edge-LDP.

\textbf{Node-CDP.} 
There have been many prior research works related to Node differential privacy (Node-DP).
For example,  a handful of graph algorithms \cite{blocki2013differentially, kasiviswanathan2013analyzing,day2016publishing, raskhodnikova2016lipschitz} have been designed for publishing the degree distribution by proposing different graph projection methods.
For instance, the truncation method~\cite{kasiviswanathan2013analyzing} removes all nodes with the degree over $\theta$. 
Edge-removal approach \cite{blocki2013differentially} traverses all edges in an arbitrary order and removes each edge connected to a node with a degree more than $\theta$.
Edge-addition method \cite{day2016publishing} traverses the edges in a stable order and inserts each edge correlated to node with degree over $\theta$.
However, the existing projection methods are only designed for Node-CDP and are not viable in Node-LDP.

\textbf{Edge-LDP.} Since there is no need for a trusted server and a large amount of valuable information resides in a decentralized social network, LDP is becoming increasingly popular in privacy protection of graph analysis. Existing works focus on various graph statistics, such as degree distribution (or histogram)\cite{LF-GDPR}, subgraph counting (e.g., k-clique, k-star, k-triangle) \cite{sun2019analyzing,imola2021locally}, synthetic graph generation \cite{qin2017generating,zhang2018two}, publishing attributed graph\cite{AsgLDP,jorgensen2016publishing}, etc. 
For instance, Ye $et$ $al.$ \cite{LF-GDPR} propose a LDP-enabled graph metric estimation framework for general graph analysis. In \cite{imola2021locally}, subgraph counting is protected locally by a more sophisticated algorithm that uses an additional round of interaction between individuals and server. 
To strike a balance between noise added to satisfy LDP and information loss from a coarser granularity, Qin $et$ $al.$ \cite{qin2017generating} design a novel multi-phase approach to synthetic decentralized social graph generation.
However, these existing works are all based on Edge-LDP which provides a weaker privacy guarantee than our work under Node-LDP.

\section{Conclusion}
\label{sec:conclusion}
To conclude, we first discuss the motivation for publishing the graph statistics under Node-LDP, and present the challenges of finishing the projection locally. 
We propose two methods for the projection parameter selection: pureLDP parameter selection and crypto-assisted parameter selection.
Also, we design two methods for executing local graph projection: node-level local projection and edge-level local projection.
Theoretical and experimental analysis verify the utility and privacy achieved by our proposed work.

\section*{Acknowledgment}
This work was partially supported by JST SPRING JPMJSP2110, JST CREST JPMJCR21M2, JST SICORP JPMJSC2107, JSPS KAKENHI Grant Numbers 21K19767, 22H03595, 22H00521.

\bibliographystyle{IEEEtran}
\bibliography{IEEEabrv,Reference}

\end{document}